\documentclass[twocolumn]{aastex61}

\definecolor{carnelian}{rgb}{0.7, 0.11, 0.11}
\definecolor{ao(english)}{rgb}{0.0, 0.5, 0.0}

\newcommand{\logne}[1]{{log$(n_\mathrm{e}\,[\mathrm{cm}^{-3}])${#1}}}
\newcommand{\logt}[1]{{log($T$\,[K]){#1}}}

\usepackage{color}
\usepackage{graphicx}

\received{\today}
\revised{}
\accepted{}
\submitjournal{ApJ}

\shorttitle{Diagnostics of $\kappa$-distributions from SDO/EVE observations}
\shortauthors{Dzif\v{c}\'{a}kov\'{a} et al.}

\begin{document}

\title{Spectroscopic diagnostics of the non-Maxwellian $\kappa$-distributions \\using SDO/EVE observations of the 2012 March 7 X-class flare}

\author{Elena Dzif\v{c}\'{a}kov\'{a}}

\author{Alena Zemanov\'{a}}

\correspondingauthor[0000-0003-1308-7427]{Jaroslav Dud\'{i}k}
\email{dudik @asu.cas.cz}

\author[0000-0003-1308-7427]{Jaroslav Dud\'{i}k}
\affiliation{Astronomical Institute of the Czech Academy of Sciences, Fri\v{c}ova 298, 251 65 Ond\v{r}ejov, Czech Republic}

\author{\v{S}imon Mackovjak}
\affiliation{Institute of Experimental Physics, Slovak Academy of Sciences, Watsonova 47, 04001 Ko\v{s}ice, Slovak Republic}

\begin{abstract}
	{Spectroscopic observations made by the Extreme Ultraviolet Variability Experiment (EVE) on board the Solar Dynamics Observatory (SDO) during the 2012 March 7 X5.4-class flare (SOL2012-03-07T00:07) are analyzed for signatures of the non-Maxwellian $\kappa$-distributions.}
	{Observed spectra were averaged over 1 minute to increase photon statistics in weaker lines and the pre-flare spectrum was subtracted. Synthetic line intensities for the $\kappa$-distributions are calculated using the KAPPA database.} 
	{We find strong departures ($\kappa$\,$\lesssim$\,2) during the early and impulsive phases of the flare, with subsequent thermalization of the flare plasma during the gradual phase. If the temperatures are diagnosed from a single line ratio, the results are strongly dependent on the value of $\kappa$. For $\kappa$\,=\,2, we find temperatures about a factor of two higher than the commonly used Maxwellian ones. The non-Maxwellian effects could also cause the temperatures diagnosed from line ratios and from the ratio of GOES X-ray channels to be different. Multithermal analysis reveals the plasma to be strongly multithermal at all times with flat DEMs. For lower $\kappa$, the DEM$_\kappa$ are shifted towards higher temperatures. The only parameter that is nearly independent of $\kappa$ is electron density, where we find \logne\,$\approx$\,11.5 almost independently of time.}
	{We conclude that the non-Maxwellian effects are important and should be taken into account when analyzing solar flare observations, including spectroscopic and imaging ones.}
\end{abstract}


%
\section{Introduction}
\label{Sect:1}

Solar flares \citep[e.g.][]{Fletcher11} are brilliant yet transient manifestations of the solar magnetic activity. During flares, magnetic reconnection \citep[e.g.,][]{Dungey53,Parker57,Sweet58,Priest00,Zweibel09,Aulanier12,Janvier13,Janvier15,Janvier17} converts excess magnetic energy into other forms, such as thermal and kinetic energies \citep[e.g.,][]{Emslie12}. A considerable portion of the released energy is converted into accelerated particles, producing enhanced high-energy tails, which are ubiquitously detected from flare free-free emission \citep[e.g.,][]{Brown71,Lin71,Holman03,Saint-Hilaire08,Krucker08b,Kasparova09b,Veronig10,Fletcher11,Holman11,Kontar11,Zharkova11,Oka13,Oka15,Simoes15,Battaglia13,Kuhar16} and occur even in microflares \citep[e.g.,][]{Hannah08,Glesener17,Wright17}. Generally, departures from the equilibrium Maxwellian distribution arise whenever particle acceleration is occurring, and the fundamental reason for existence of high-energy tails is the $\sim$$E^{-2}$ behavior of the electron collisional cross-section with the kinetic energy $E$ \citep{Scudder79,Meyer-Vernet07,Scudder13}. 

In this work, we study the influence of the high-energy tails on the intensities of the optically thin emission lines produced at flare temperatures. To quantify the departure from Maxwellian, we utilize the non-Maxwellian $\kappa$-distributions. These distributions are characterized by a power-law high-energy tail, and occur naturally in situations characterized by turbulence \citep{Hasegawa85,Laming07} which happen under flare conditions as well \citep{Bian14}. Indeed, indications of the $\kappa$-distributions have been obtained from flare observations. \citet{Kasparova09b} showed that the bremsstrahlung spectra arising from flare plasma at coronal altitudes can be described by a $\kappa$-distribution. In their event, the flare chromospheric footpoint emission was occulted by the solar limb. \citet{Oka13,Oka15} showed that the $\kappa$-distributions provide a good description of the high-energy tail detected in above-the-loop-top sources, although a thermal Maxwellian component was also present.
Indications of $\kappa$-distributions of ions with extremely non-Maxwellian values of $\kappa$ were also found by \citet{Jeffrey16,Jeffrey17}. These authors studied the emission arising in flare loop-top, ribbon, and hard X-ray footpoints, and showed that the emission line profiles are well-described by a $\kappa$-distribution.

Indications of the electron $\kappa$-distributions were also found in a transient coronal loop occurring in the same location as a previous B-class flare \citep{Dudik15}. These authors analyzed the \ion{Fe}{11}--\ion{Fe}{12} emission line ratios observed by the Extreme-Ultraviolet Imaging Spectrometer \citep[EIS,][]{Culhane07} onboard the Hinode satellite. The \ion{Fe}{11}--\ion{Fe}{12} line ratios were found to be strongly non-Maxwellian. Indications of the strongly non-Maxwellian distributions of both electrons and ions were also found in the transition region observations performed by the Interface Region Imaging Spectrograph \citep{DePontieu14}. Indications of the non-Maxwellian ions were found from the line profiles of \ion{Si}{4} and \ion{O}{4}, and electrons from the relative intensities of these lines \citep{Dudik17b}. The values of $\kappa$ found from line profiles and line intensities were similar.

A review of the applications of the $\kappa$-distributions in solar physics can be found in \citet{Dudik17a}. Other astrophysical applications can be found e.g. in \citet{Pierrard10} and \citet{Bykov13}. Finally we note that the $\kappa$-distributions can be used for description of plasma with multiple Maxwellian components \citep{Hahn15b}. Most of the additional Maxwellians are used to approximate the tail of the distribution, and their relative amplitudes decrease with increasing temperatures of these Maxwellians. In principle, a $\kappa$-distribution could thus represent a special case of multi-thermal plasma. \citet{Battaglia15} used a differential emission measure (DEM) represented by a $\kappa$-distribution and fitted it to observations of a single-loop flare performed simultaneously by the Reuven-Ramaty High-Energy Solar Spectroscopic Imager \citep[RHESSI,][]{Lin02} and the Atmospheric Imaging Assembly \citep[AIA,][]{Lemen12,Boerner12} onboard the Solar Dynamics Observatory \citep[SDO,][]{Pesnell12}. This analysis yielded $\kappa$\,$\approx$\,4.

This paper is organized as follows. The flare selected for analysis (SOL2012-03-07T00:07) and its observations by the Extreme Ultraviolet Variability Experiment \citep[EVE,][]{Woods12} onboard SDO are described in Sect. \ref{Sect:2}. The synthesis of non-Maxwellian optically thin spectra are detailed in Sect. \ref{Sect:3}. In Sect. \ref{Sect:4}, we describe the diagnostics of the flare plasma, including diagnostics of electron density (Sect. \ref{Sect:4.1}), temperature (Sect. \ref{Sect:4.2}), the parameter $\kappa$ (Sect. \ref{Sect:4.3}), differential emission measure (Sect. \ref{Sect:4.4}), as well as its influence on diagnostics of $\kappa$ (Sect. \ref{Sect:4.5}). A summary of the results is given in Sect. \ref{Sect:5}. Details on EVE lines and their blends are given in Appendix \ref{Appendix:A}.

%
%
\begin{figure*}[ht]
	\centering 
	\includegraphics[width=8.8cm,clip]{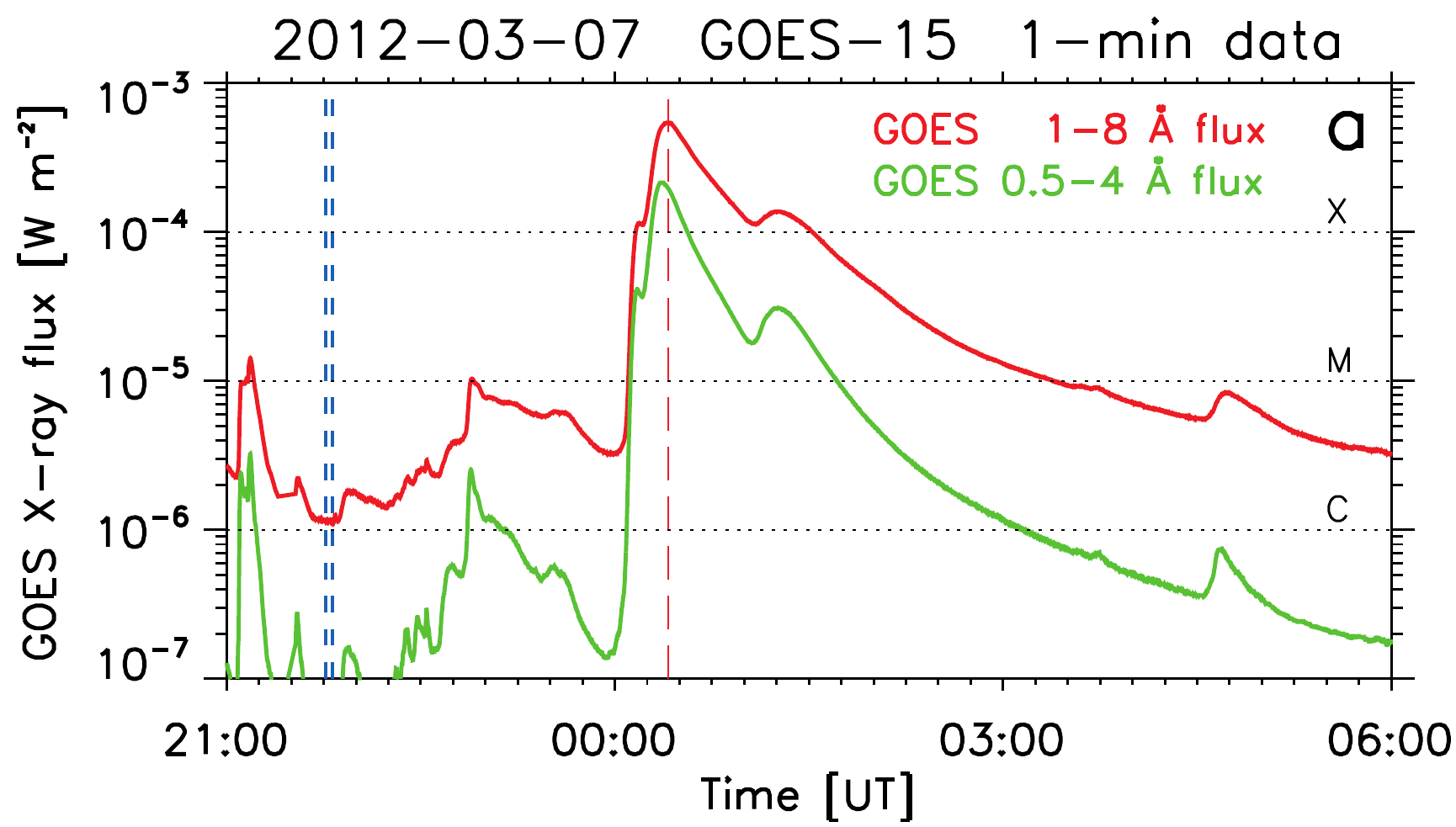}
	\includegraphics[width=8.8cm,clip]{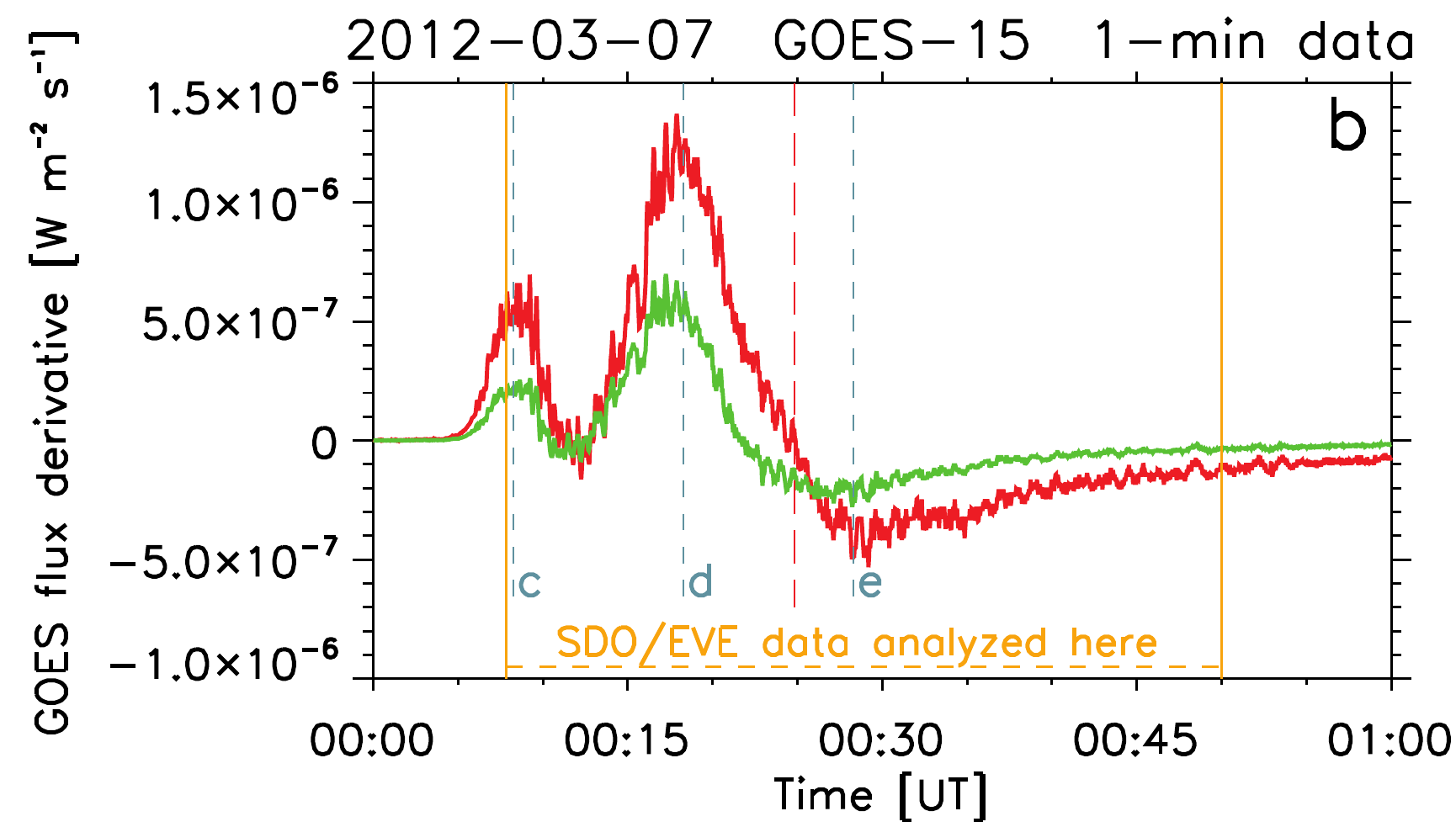}
	\includegraphics[width=6.63cm,bb= 0 0 490 425,clip]{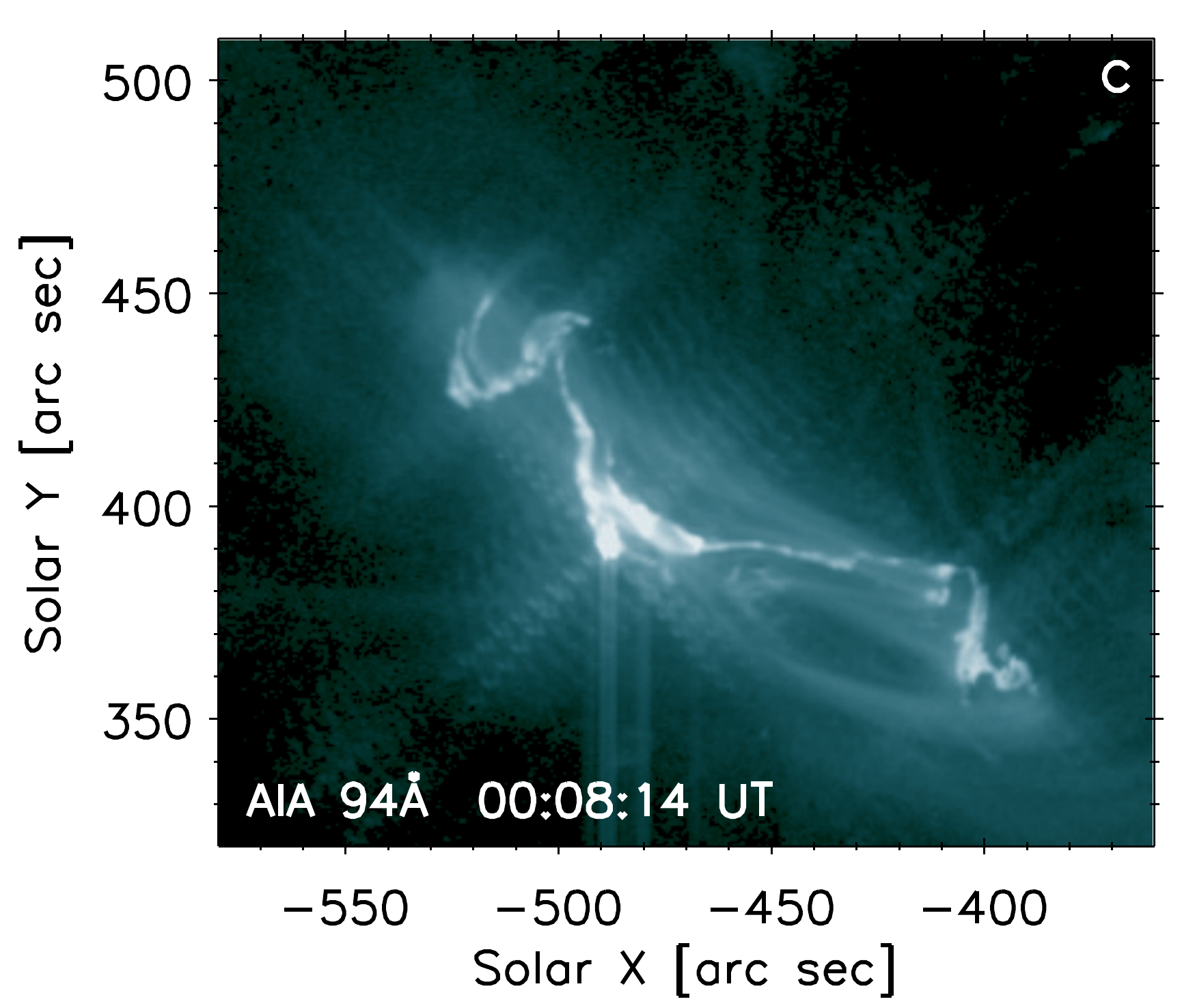}
	\includegraphics[width=5.48cm,bb=85 0 490 425,clip]{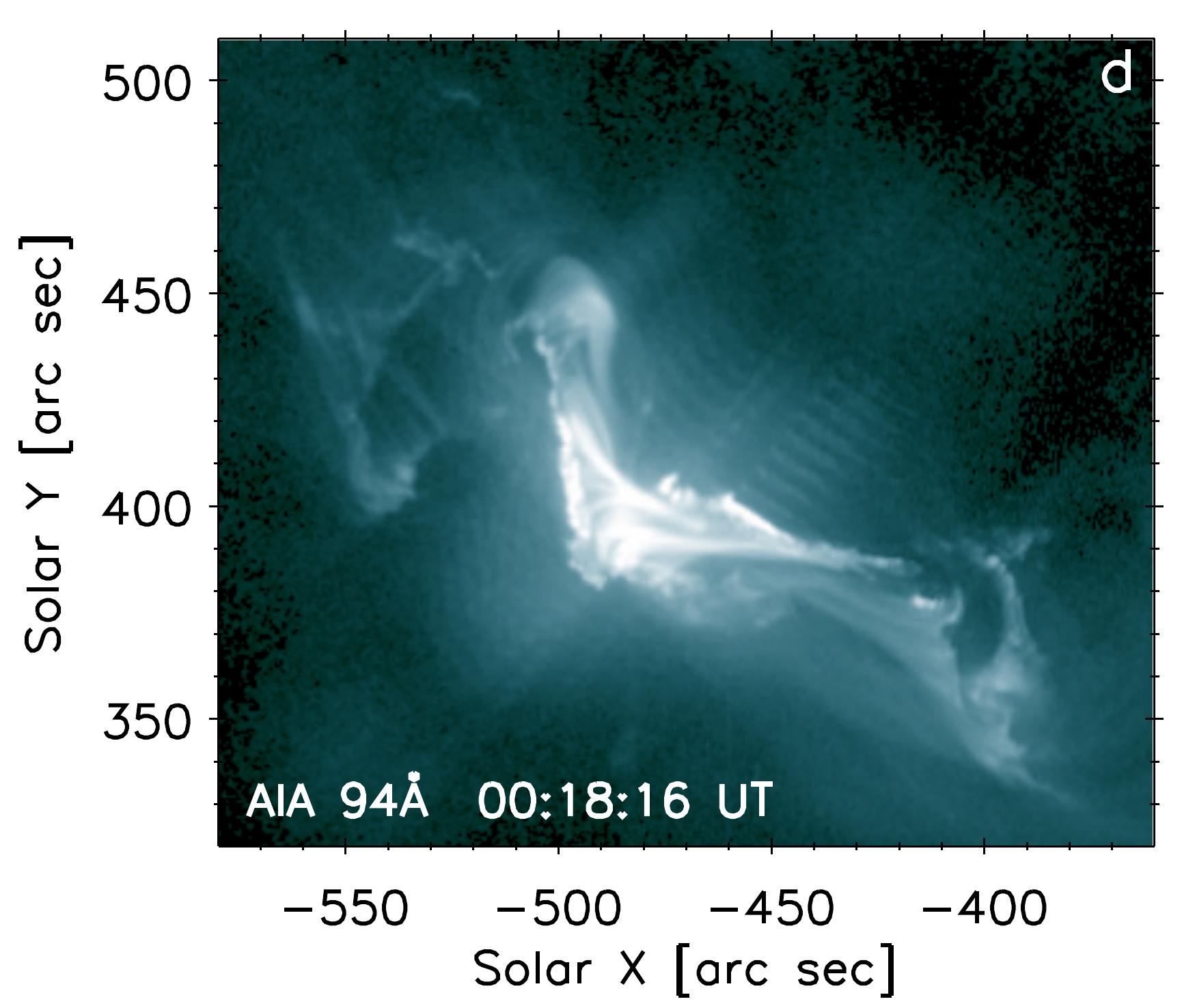}
	\includegraphics[width=5.48cm,bb=85 0 490 425,clip]{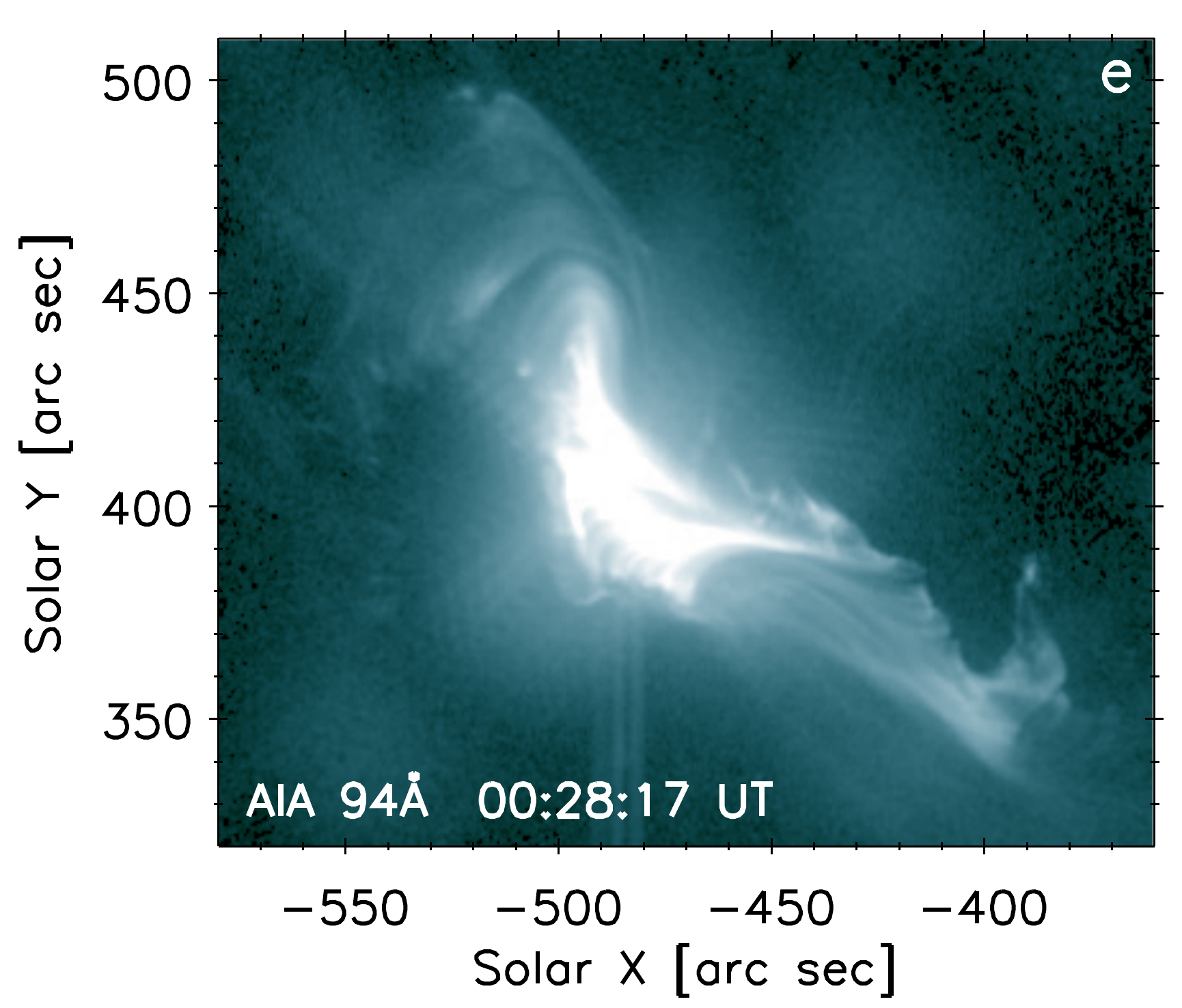}
\caption{Context observations of the X5.4-class flare of 2012 March 7 from GOES-15 and SDO/AIA. (a)--(b): GOES-15 X-ray flux and its derivative, respectively. Red and blue vertical dashed lines indicate the time of the flare peak and the interval where the pre-flare spectrum was selected, respectively. (c)--(e): SDO/AIA imaging observations (log scaled) of the flare in the 94\,\AA~channel dominated by \ion{Fe}{18}. The times shown are indicated in panel (b) by green vertical dashed lines.}
\label{Fig:Context}
\end{figure*}
%
%
%
\section{Observations}
\label{Sect:2}

%
\subsection{The X5.4-class flare of 2012 March 07}
\label{Sect:2.1}

The X5.4-class flare of 2012 March 07 (SOL2012-03-07T00:07) is the fourth largest flare of the current Solar cycle 24, according to the X-ray Flare Dataset\footnote{https://www.ngdc.noaa.gov/stp/space-weather/solar-data/solar-features/solar-flares/x-rays/goes/xrs/}. It occurred in the Active region NOAA 11429, which was a well-known flaring region studied by many authors \citep[e.g.,][]{Doschek13,Simoes13a,Schrijver15,Brown16,Harra16,Polito17,Dudik17c}. On 2012 March 07, the AR 11429 possessed a $\delta$-spot in anti-Hale configuration, a situation prone to strong flaring \citep{Chintzoglou15}. The X5.4-class flare was followed in its gradual phase by another X1.3-class flare about an hour later (Fig. \ref{Fig:Context}). These two flares were sources of two super-fast CMEs \citep{Chintzoglou15}.

Spectroscopic analysis of the Hinode/EIS observations of confined flares occurring prior to the eruptive ones were performed by \citet{Syntelis16}. Formation of the two erupting flux ropes during the confined flares and the pre-eruptive magnetic geometry were studied by \citet{Chintzoglou15}. The hydrogen Lyman series and \ion{C}{3} emission of the X-class flare from SDO/EVE was examined by \citet{Brown16}. Other aspects of the X-class flares, such as the $\gamma$-ray and proton observations, as well as the CMEs and their propagation were studied by \citet{Ajello14}, \citet{Kouloumvakos16} and \citet{Patsourakos16}. The flare was not observed by RHESSI, which started its observations only after 02:05\,UT.

The X-ray flux and its derivative during the X5.4-class flare are shown in Fig. \ref{Fig:Context}, panels (a) and (b), respectively. A strong rise of the X-ray flux started at about 00:06 UT on 2012 March 07, followed by the impulsive phase. The flare reached its maximum at about 00:25\,UT (red dashed line in Fig. \ref{Fig:Context}a) and progressed to the gradual phase. The morphology of the flare is shown in panels (c) to (e) of Fig. \ref{Fig:Context}. There, the imaging observations performed by the SDO/AIA instrument \citep{Lemen12,Boerner12} in its 94\,\AA~channel are shown. The 94\,\AA~channel is dominated by \ion{Fe}{18} 93.93\,\AA~emission under flaring conditions \citep{ODwyer10,Petkaki12}. The morphology at the flare onset (panel c) is suggestive of a sheared magnetic configuration \citep[c.f.,][]{Chintzoglou15} with brightenings close to the polarity inversion line. Subsequently, the flare develops into an arcade of flare loops, growing both laterally along the polarity inversion line as well as across it, with decreasing magnetic shear, in agreement with the Standard solar flare model in 3D \citep{Aulanier12,Janvier13,Janvier15}. The flare is very bright and thus most of the AIA pass-bands are saturated even at lower exposure times.

%
\subsection{SDO/EVE observations of the flare}
\label{Sect:2.2}

The Extreme Ultraviolet Variability Experiment \citep[EVE,][]{Woods12} on board the Solar Dynamics Observatory \citep[SDO,][]{Pesnell12} is a collection of instruments for measuring the solar EUV irradiance from 1 to 1050\,\AA~with spectral resolution of $\approx$1\,\AA~at a cadence of about 10\,s. For our purposes, we used the data obtained by the Multiple EUV Grating Spectrographs A and B. The MEGS-A was a routinely operating (until 2014 May 26) grazing-incidence, off-Rowland circle spectrograph measuring at 50--370\,\AA. The MEGS-B is a normal-incidence, dual-pass spectrograph operating at wavelengths above 350\AA~and up to 1050\,\AA. The MEGS-B instrument suffered degradation limiting its operations.

Both MEGS-A and B instruments observed the 2012 March 07 flare at full cadence of 10\,s throughout the rise, impulsive, peak, and gradual phases of the flare. Here, we analyze both MEGS-A and B observations made during 00:08 -- 00:50 UT. During this interval, the flare lines, including the weaker lines required for diagnostics (Sect. \ref{Sect:4}) are well-observed. This time interval captures nearly the entirety of the flare from the early phase up to the beginning of its gradual phase (c.f., Fig. \ref{Fig:Context}b).

The EVE observations of the flare were analyzed by \citet{DelZanna13c}. There, example spectra during the pre-flare, impulsive, peak, and gradual phases are shown together with the lightcurves of the selected strong lines, especially Fe lines from various ionization stages \citep[\ion{Fe}{9}--\ion{Fe}{23}; see also][]{Harra16}. Diagnostics of temperature, electron density, and emission measure were also performed and discussed by \citet{DelZanna13c}. The low EVE spectral resolution of $\approx$1\,\AA~means that most of the lines observed are blended. The known and unknown blends, their wavelengths, contribution to the intensity of the main line, behavior with temperature and flare evolution were also discussed by \citet{DelZanna13c}. 

%
%
\section{Non-Maxwellian line intensity calculations}
\label{Sect:3}

Here, we re-visit the EVE flare observations to perform non-Maxwellian diagnostics of the plasma, as well as to analyze the influence of the departures from the Maxwellian on the diagnosed temperature $T$ and electron density $n_\mathrm{e}$. To do that, we use the non-Maxwellian $\kappa$-distributions \citep[e.g.,][]{Olbert68,Vasyliunas68a,Vasyliunas68b,Owocki83,Livadiotis15b,Dzifcakova15}
\begin{equation}
	f_\kappa(E)dE= A_\kappa  \frac{2}{\sqrt{\pi}\left(k_\mathrm{B}T_\kappa\right)^{3/2}} \frac{E^{1/2}dE}{\left(1+\frac{E}{(\kappa-3/2)k_\mathrm{B}T_\kappa}\right)^{\kappa+1}}\,,
 \label{Eq:Kappa}
\end{equation}
which allows for modeling of the effect of the high-energy tails by using only one extra free parameter, $\kappa$. Maxwellian distribution is recovered for $\kappa$\,$\to$\,$\infty$, while extreme non-Maxwellian situations occur for $\kappa$\,$\to$\,3/2. In Eq. (\ref{Eq:Kappa}), $E$ is the electron kinetic energy,  $k_\mathrm{B}$\,=\,1.38\,$\times$10$^{-16}$ erg\,K$^{-1}$ is the Boltzmann constant, and $A_\kappa$\,=\,$\Gamma(\kappa+1) / (\Gamma(\kappa-1/2) (\kappa-3/2)^{3/2}$ is the normalization constant.

The $\kappa$-distribution is characterized by a near-Maxwellian core with temperature $T_\mathrm{C}$\,=\,$(\kappa-3/2)/\kappa$ \citep[][Section 2 and Figure 1 therein]{Oka13} and a power-law high-energy tail with the power-law index of $\kappa+1/2$ (Eq. \ref{Eq:Kappa}). In terms of the power-law index of bremsstrahlung radiation \citep[see also][]{Dudik12} routinely observed in X-rays in case of a thin-target source, $\gamma_\mathrm{thin}$\,=\,$\delta$+1\,=\,$\delta'$+1/2, where $\gamma_\mathrm{thin}$ is the power-law index of the photon flux spectrum, and $\delta$ and $\delta'$ are the power-law indices of electron energy flux and energy distributions, respectively \citep[c.f.,][]{Brown71,Tandberg88}. For a $\kappa$-distribution, $\delta'$\,=\,$\kappa$+1/2 (Equation \ref{Eq:Kappa}). This means that $\delta$\,=\,$\kappa$ and $\gamma_\mathrm{thin}$\,=\,$\kappa+1$ \citep[see][]{Dudik12,Dudik17a}.

The behavior of the emission lines with $\kappa$ is more complicated. The intensity $I_{ji}$ of a spectral line arising from plasma along a line of sight $l$ is given by  \citep[cf.,][]{Mason94,Phillips08}
\begin{equation}
	I_{ji} = \int A_X G_{X,ji}(T,n_\mathrm{e},\kappa) n_\mathrm{e} n_\mathrm{H} \mathrm{d}l\,,
	\label{Eq:line_intensity}
\end{equation}
where $G_{X,ji}(T,n_\mathrm{e},\kappa)$ is the line contribution function
\begin{equation}
	G_{X,ji}(T,n_\mathrm{e},\kappa) = \frac{hc}{\lambda_{ji}} \frac{A_{ji}}{n_\mathrm{e}} \frac{n(X_j^{+k})}{n(X^{+k})} \frac{n(X^{+k})}{n(X)}\,.
	\label{Eq:G(T)}
\end{equation}
In these equations, $j$ and $i$ stand for the upper and lower level corresponding to the radiative transition arising from the $X^{+k}$ ion of the element $X$ of abundance $A(X)$. The corresponding wavelength is denoted as $\lambda_{ji}$ and the Einstein coefficient as $A_{ji}$. The fractions $n(X_j^{+k}) / n(X^{+k})$ and $n(X^{+k})/n(X)$ denote the density of excited fraction of the ion $X^{+k}$ and the relative abundance of this ion, respectively. These ratios are both a function of $\kappa$ due to the dependence of individual excitation, deexcitation, ionization, and recombination rates on $\kappa$ \citep[e.g.,][]{Dzifcakova92,Dzifcakova02,Dzifcakova13a,Dudik14b,Dzifcakova15}. These rates are integral quantities of the respective cross-sections over the electron energy distribution. Thus, collisional processes across many orders of electron energies are involved in the line intensity calculation. The rates of all processes show significant departures from the Maxwellian with decreasing $\kappa$. For small $\kappa$, the ionization and ionization rates are increased by orders of magnitude at low $T$ \citep[e.g.,][]{Dzifcakova06,Dzifcakova13a,Dudik14a} compared to Maxwellian. The recombination rates are increased by a factor of about two for $\kappa$\,=\,2, however, the peak of the dielectronic recombination can be shifted to higher $T$.

In inhomogeneous situations involving many emitting structures along a given line of sight, or in case of EVE indeed the full Sun, the expression (\ref{Eq:line_intensity}) is usually recast as
\begin{equation}
 	I_{ji} = \int A_X G_{X,ji}(T,n_\mathrm{e},\kappa) \mathrm{DEM}_\kappa (T) \mathrm{d}T\,,
	\label{Eq:line_intensity_DEM}
\end{equation}
where the quantity DEM$_\kappa (T) = n_\mathrm{e} n_\mathrm{H} \mathrm{d}l / \mathrm{d}T$ is the differential emission measure, i.e., the contribution to total emission measure along the line of sight from plasma at a given $T$. Here, the subscript $\kappa$ indicates that the DEM can be a function of $\kappa$ \citep[c.f.,][]{Mackovjak14,Dudik15}.

Spectral synthesis and calculation of line intensities for the $\kappa$-distributions were performed using the KAPPA\footnote{http://kappa.asu.cas.cz} database \citep{Dzifcakova15}. KAPPA is based on the CHIANTI database and software, version 7.1 \citep{Dere97,Landi13}. We note that CHIANTI has been updated to version 8 \citep{DelZanna15b}; however, the atomic data for the \ion{Fe}{18}--\ion{Fe}{23} that we use here are the same in CHIANTI versions 7.1 and 8. The main atomic data used for level population were obtained by \citet{Witthoeft06} and \citet{DelZanna06} for \ion{Fe}{18}, \citep{Gu03} and \citet{Landi06} for \ion{Fe}{19}, \citet{Witthoeft07} for \ion{Fe}{20}, \citet{Badnell01a} and \citet{Landi06} for \ion{Fe}{21}, \citet{Badnell01b} and \citet{Landi06} for \ion{Fe}{22}, and \citet{Chidichimo05} and \citet{DelZanna05} for \ion{Fe}{23}. The former references listed stand for the effective collision strengths, while the latter for the $A$-values. Finally, for \ion{Fe}{24}, we use the atomic data of \citet{Berrington97} and \citet{Whiteford02} available within CHIANTI v7.1 and KAPPA databases. These are different from the atomic data available within CHIANTI v8, which relies on \citet{Whiteford01} and \citet{Badnell11}. The different atomic datasets result in very similar intensities; the difference for typical flare conditions is about 14\% for the 192.03\,\AA~line used here.

Finally, atomic data for ionization and recombination used for ionization equilibrium calculations \citep{Dzifcakova13a} were taken from the works of \citet{Dere07} and \citet{Dere09} for ionization, and \citet{Badnell03}, \citet{Colgan03}, \citet{Colgan04}, \citet{Mitnik04}, \citet{Badnell06a}, \citet{Altun05,Altun06,Altun07},  \citet{Zatsarinny05a}, \citet{Zatsarinny05b}, \citet{Zatsarinny06}, \citet{Bautista07}, and \citet{Nikolic10} for recombination.

%
%
\begin{figure}
	\centering
	\includegraphics[width=7.5cm,clip]{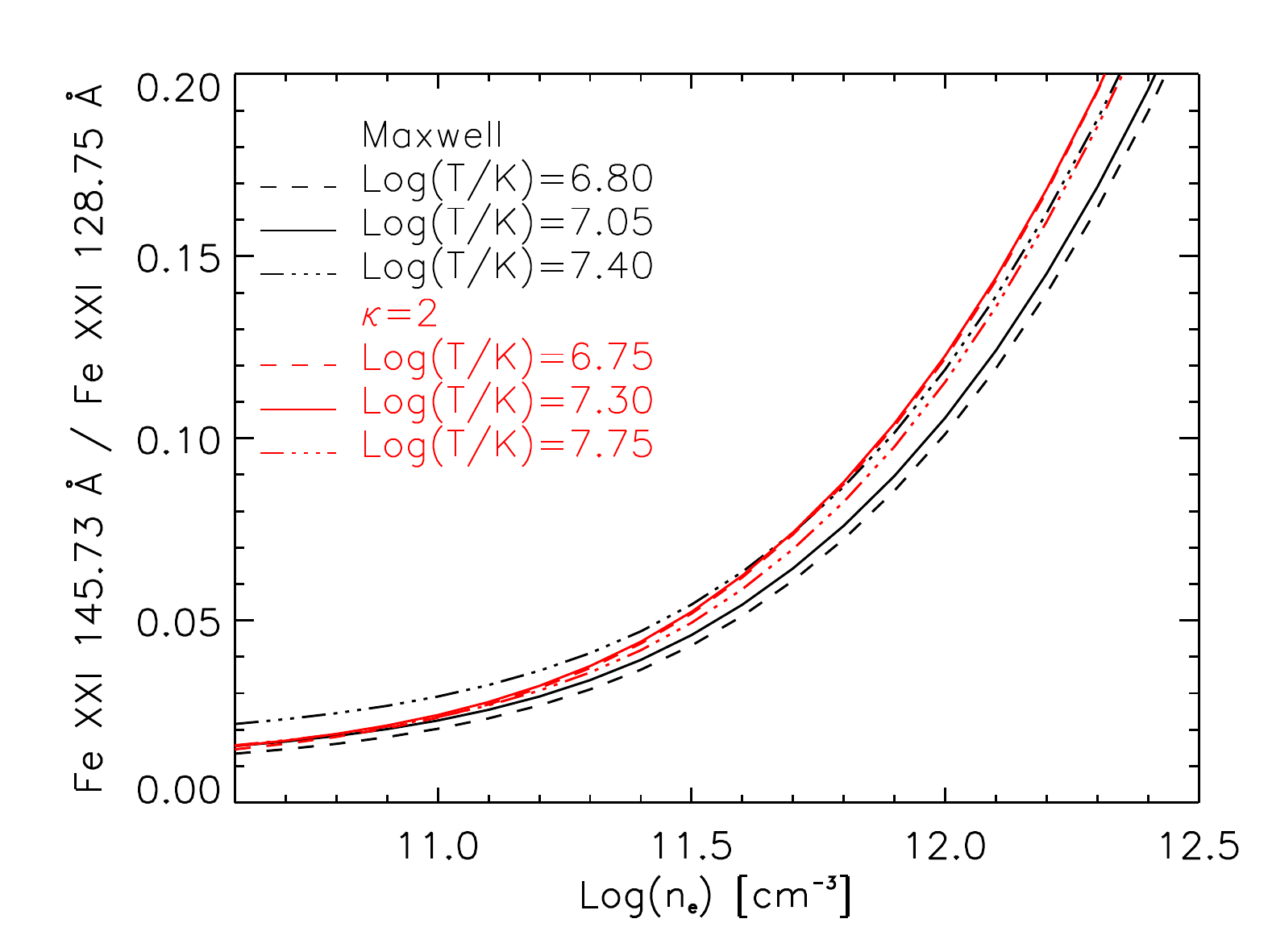}
	\includegraphics[width=7.5cm,clip]{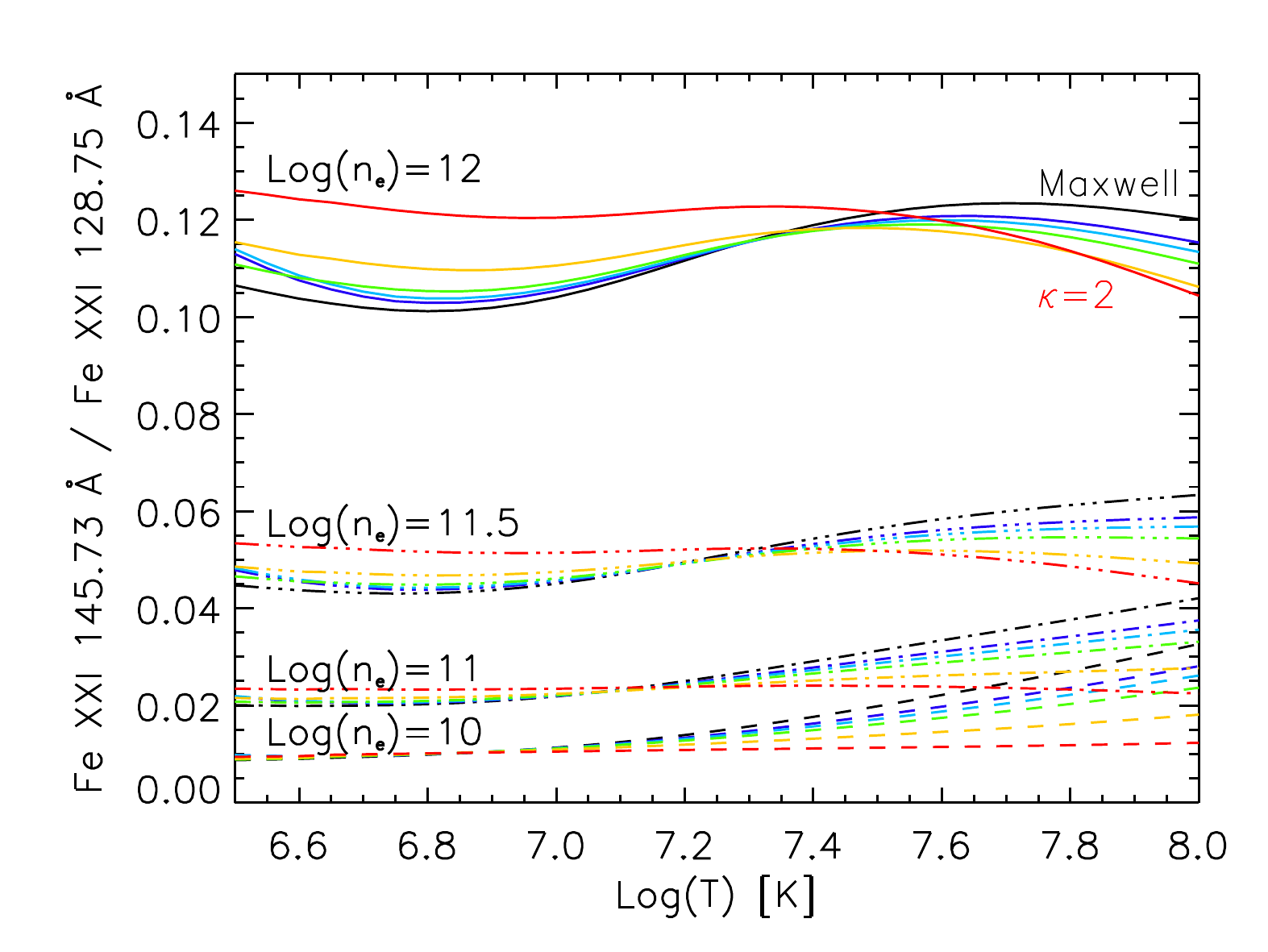}
	\includegraphics[width=7.5cm,clip]{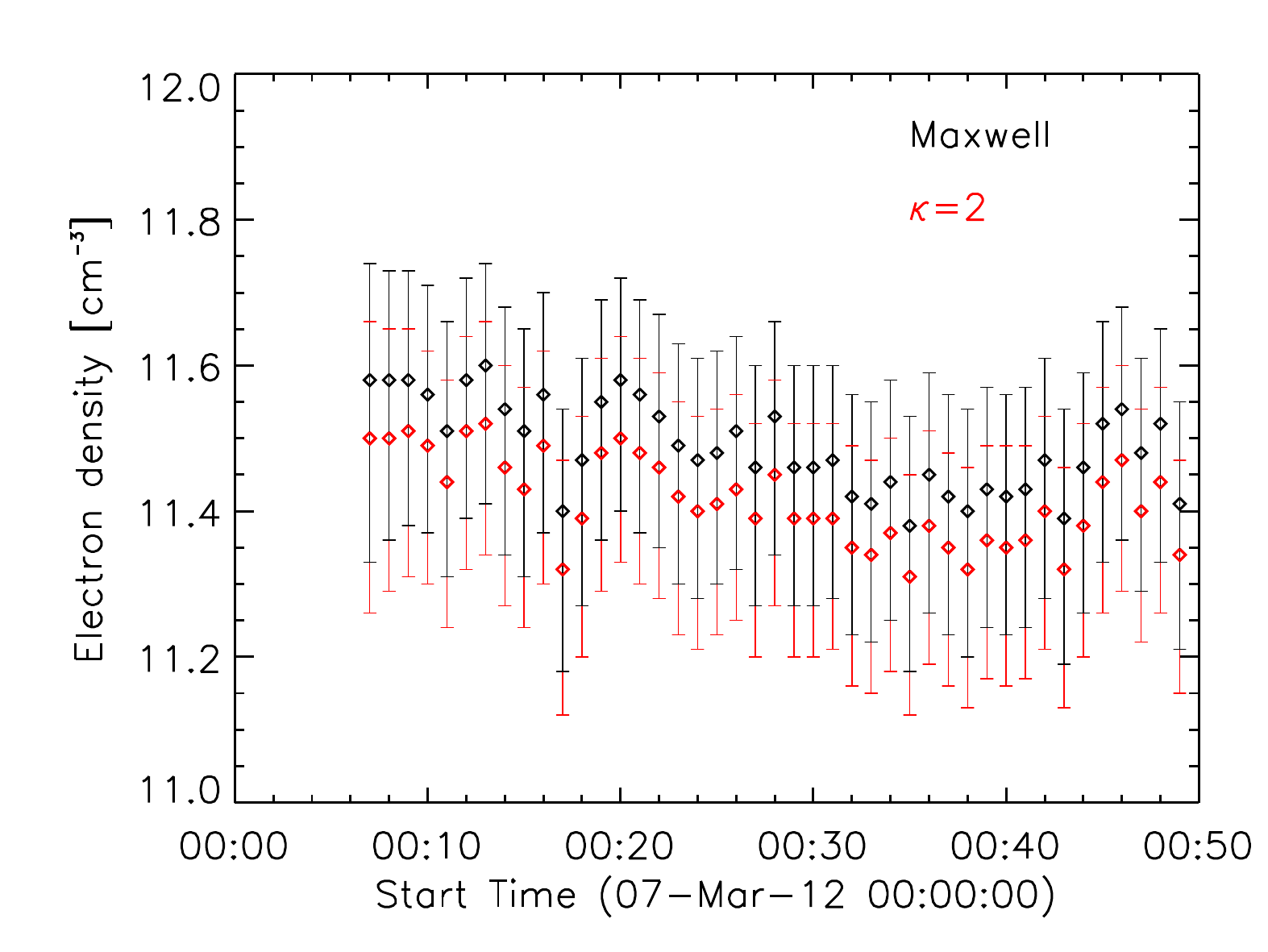}
\caption{Electron density diagnostics. \textit{Top} and \textit{middle}: Theoretical dependence of the \ion{Fe}{21} 145.73\,\AA\,/\,128.75\,\AA~ratio on electron density as a function of $T$ and $\kappa$. Colors stand for different $\kappa$, while individual linestyles denote dependence on $T$ (\textit{top}) or $n_\mathrm{e}$ (\textit{middle}). \textit{Bottom}: Diagnosed values of $n_\mathrm{e}$ and their uncertainties as a function of time, assuming either a Maxwellian (black) or $\kappa$\,=\,2 (red) distribution.}
\label{Fig:Diag_ne}
\end{figure}

%
\begin{figure*}
	\centering
	\includegraphics[width=17.5cm,clip]{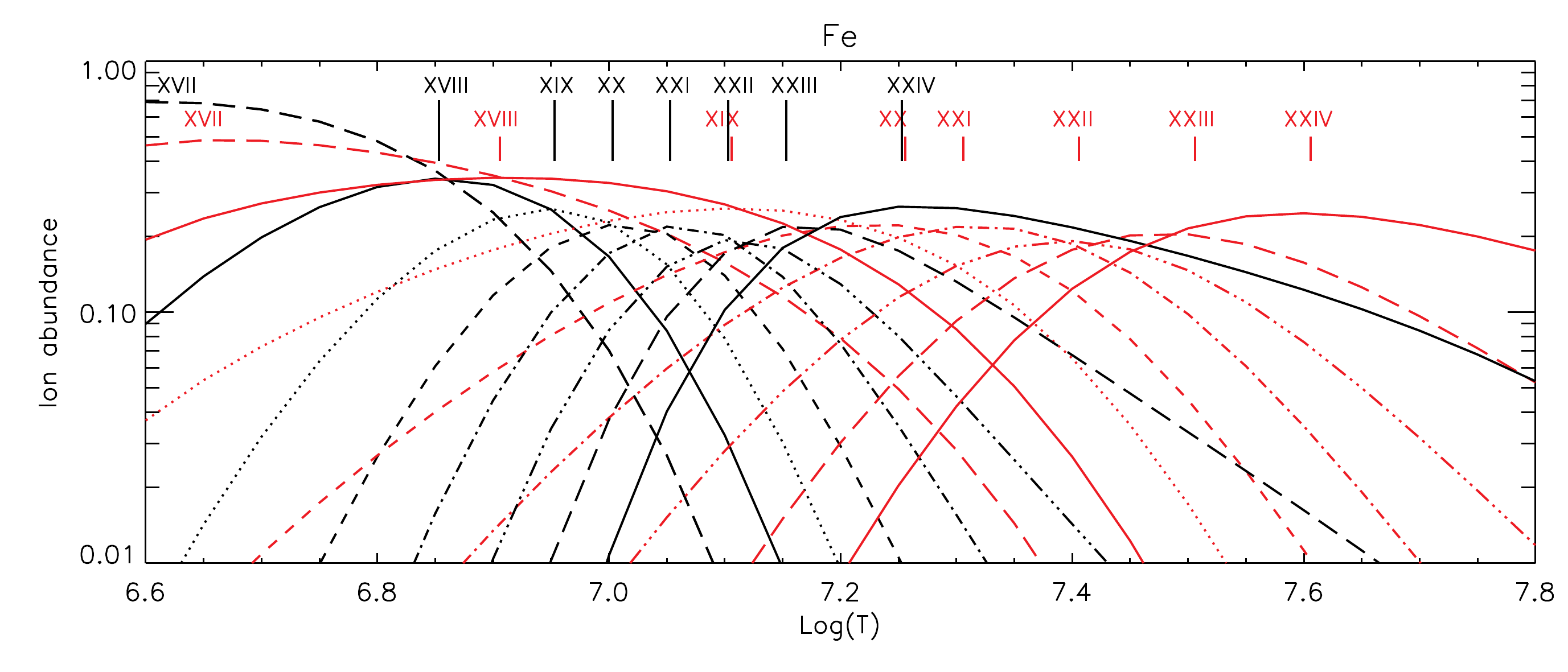}
\caption{Relative ion abundances in ionization equilibrium as a function of $T$ plotted for the Maxwellian distribution (black) and $\kappa$\,=\,2 (red color). Loci of the peaks of the abundance of individual ions are shown by vertical lines with labels.}
\label{Fig:Ioneq}
\end{figure*}  

%
\begin{figure}
	\centering
	\includegraphics[width=7.5cm,clip]{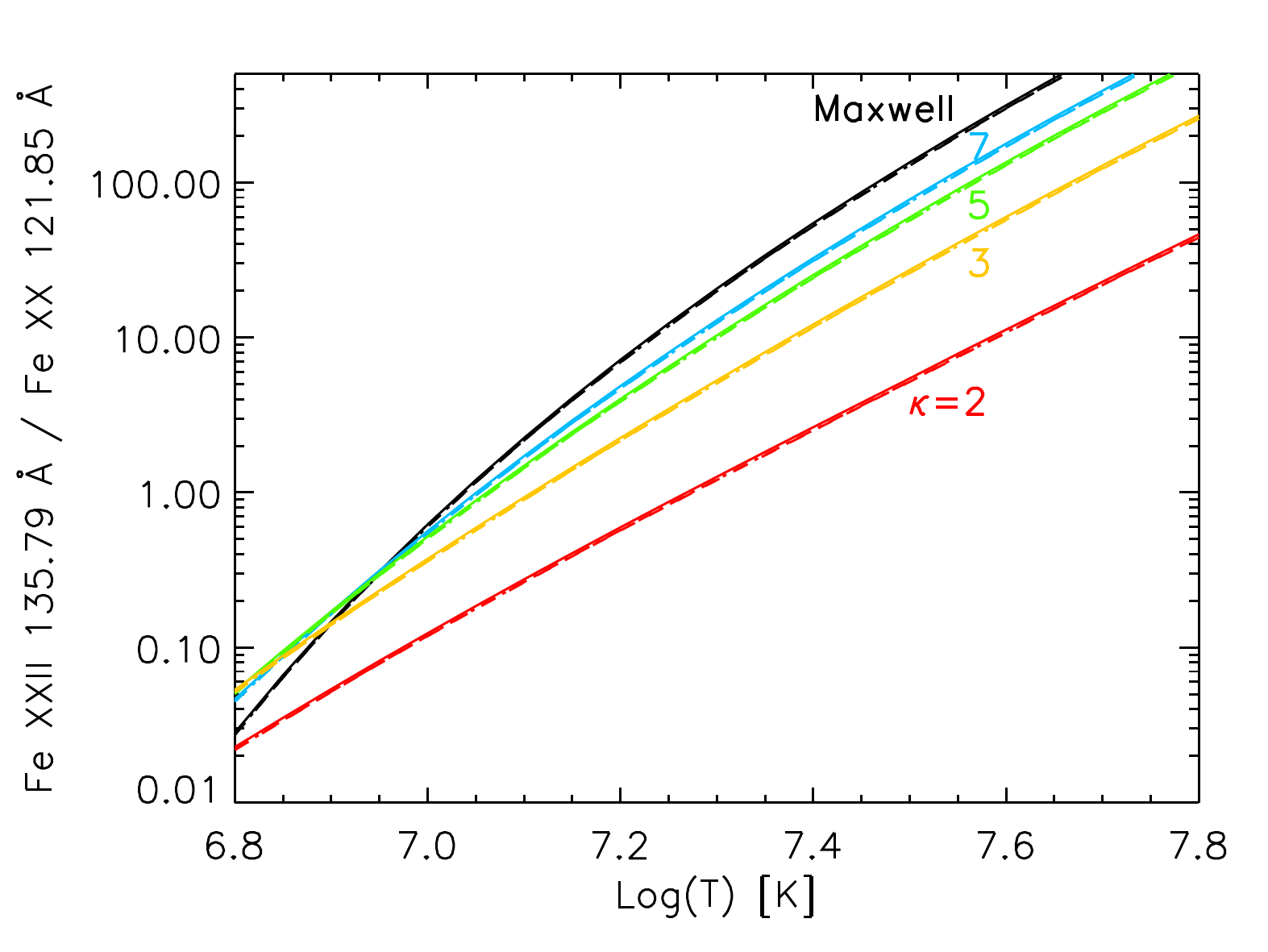}
	\includegraphics[width=7.5cm,clip]{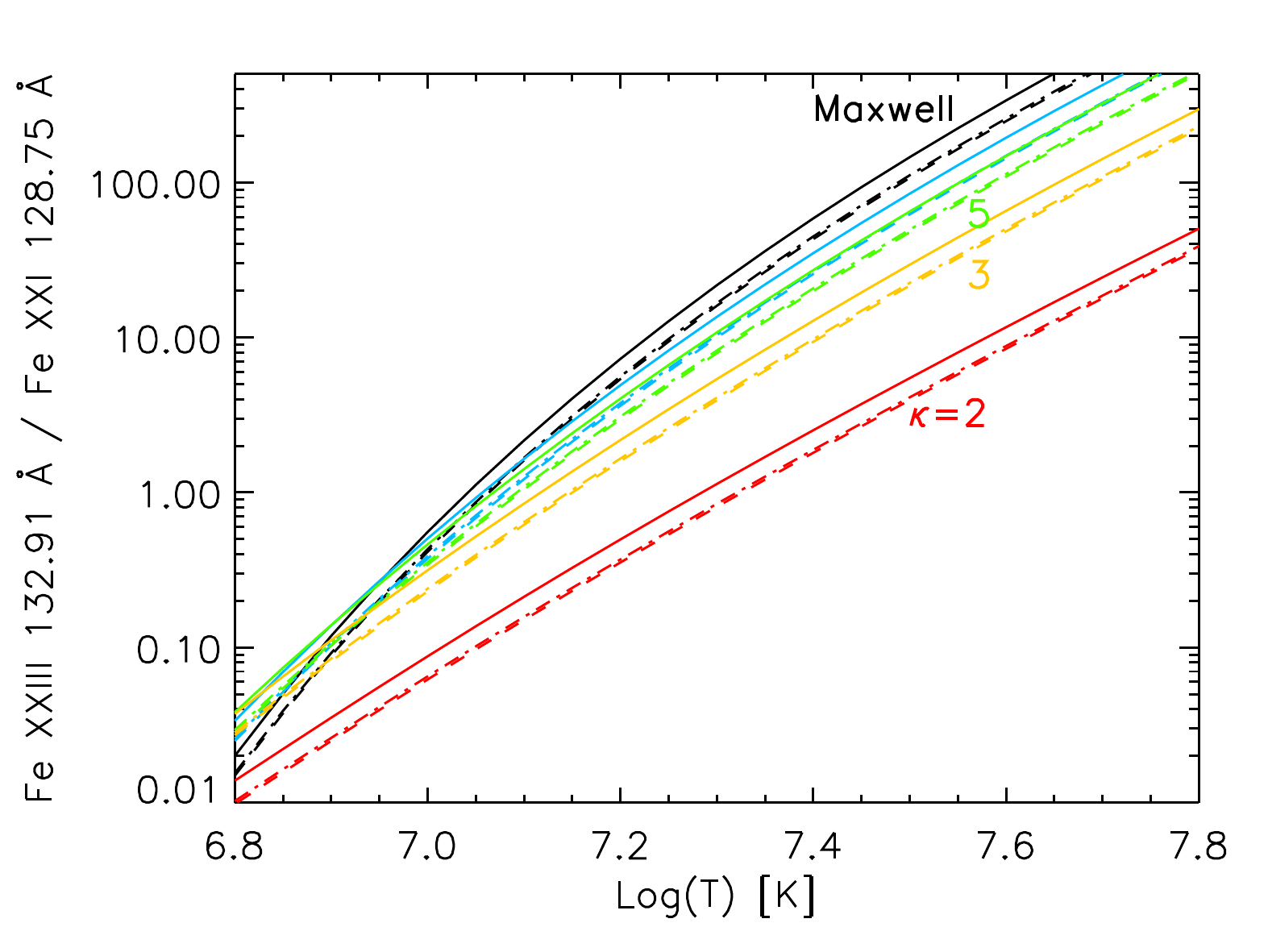}
	\includegraphics[width=7.5cm,clip]{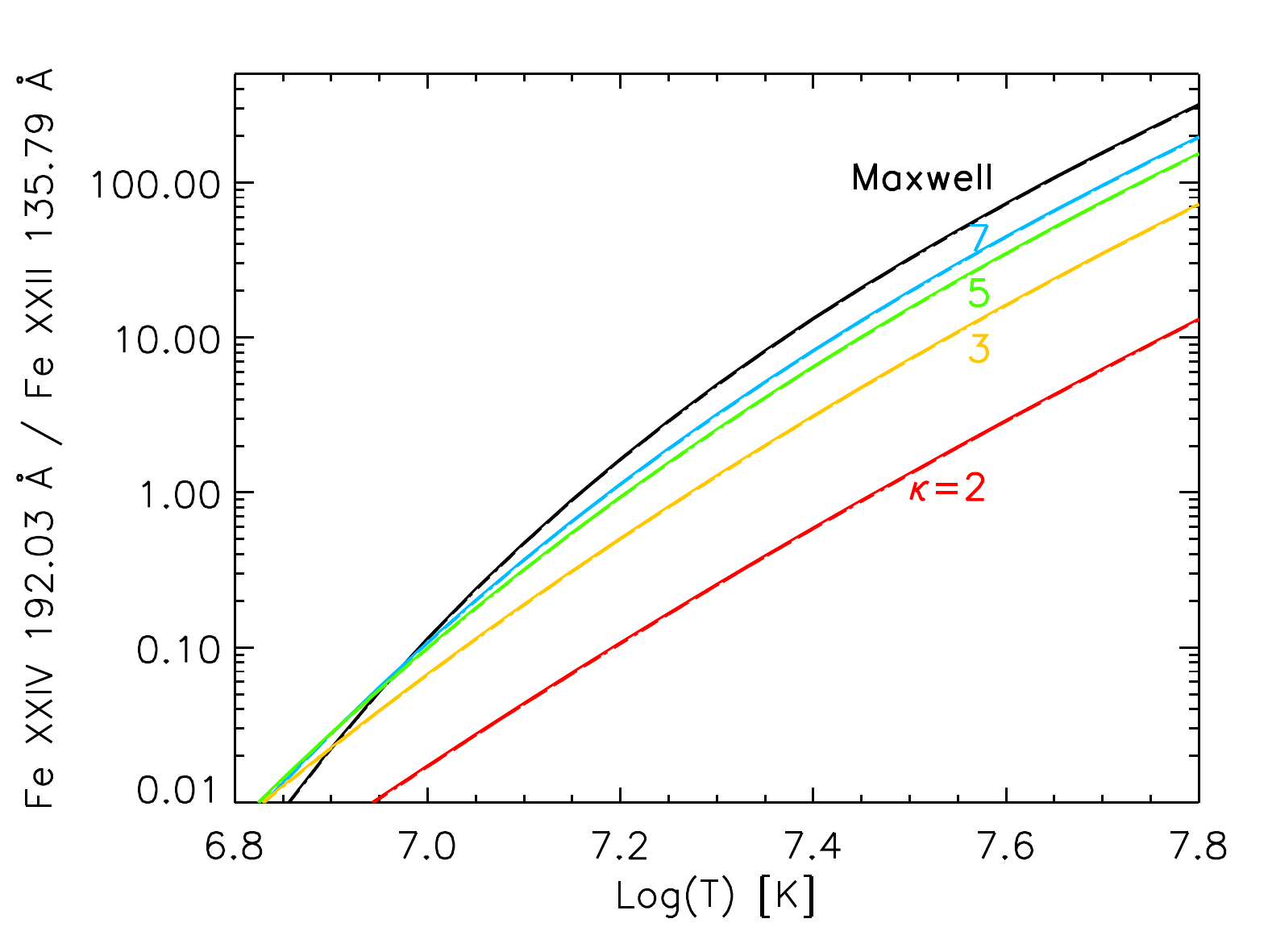}
\caption{Temperature diagnostics from line intensity ratios as a function of $\kappa$. Colors denote individual values of $\kappa$. Individual linestyles denote different electron densities. Full and dashed lines stand for log($n_\mathrm{e}$\,[cm$^{-3}$])\,=\,11 and 12, respectively. Note the logarithmic scaling of the line intensity ratio axes.}
\label{Fig:Diag_T_theor}
\end{figure}

%
\begin{figure}
	\centering
	\includegraphics[width=7.5cm,clip]{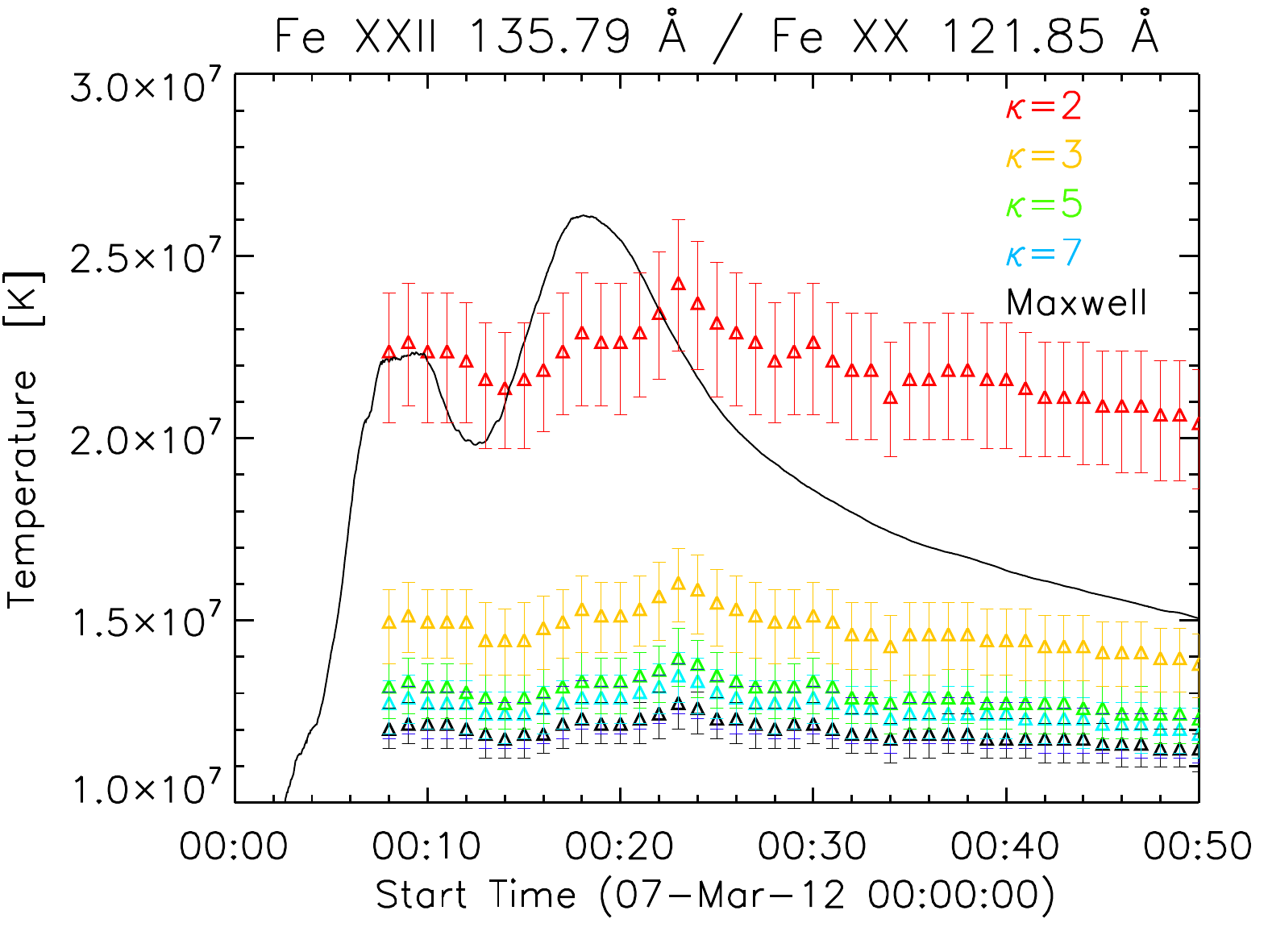}
	\includegraphics[width=7.5cm,clip]{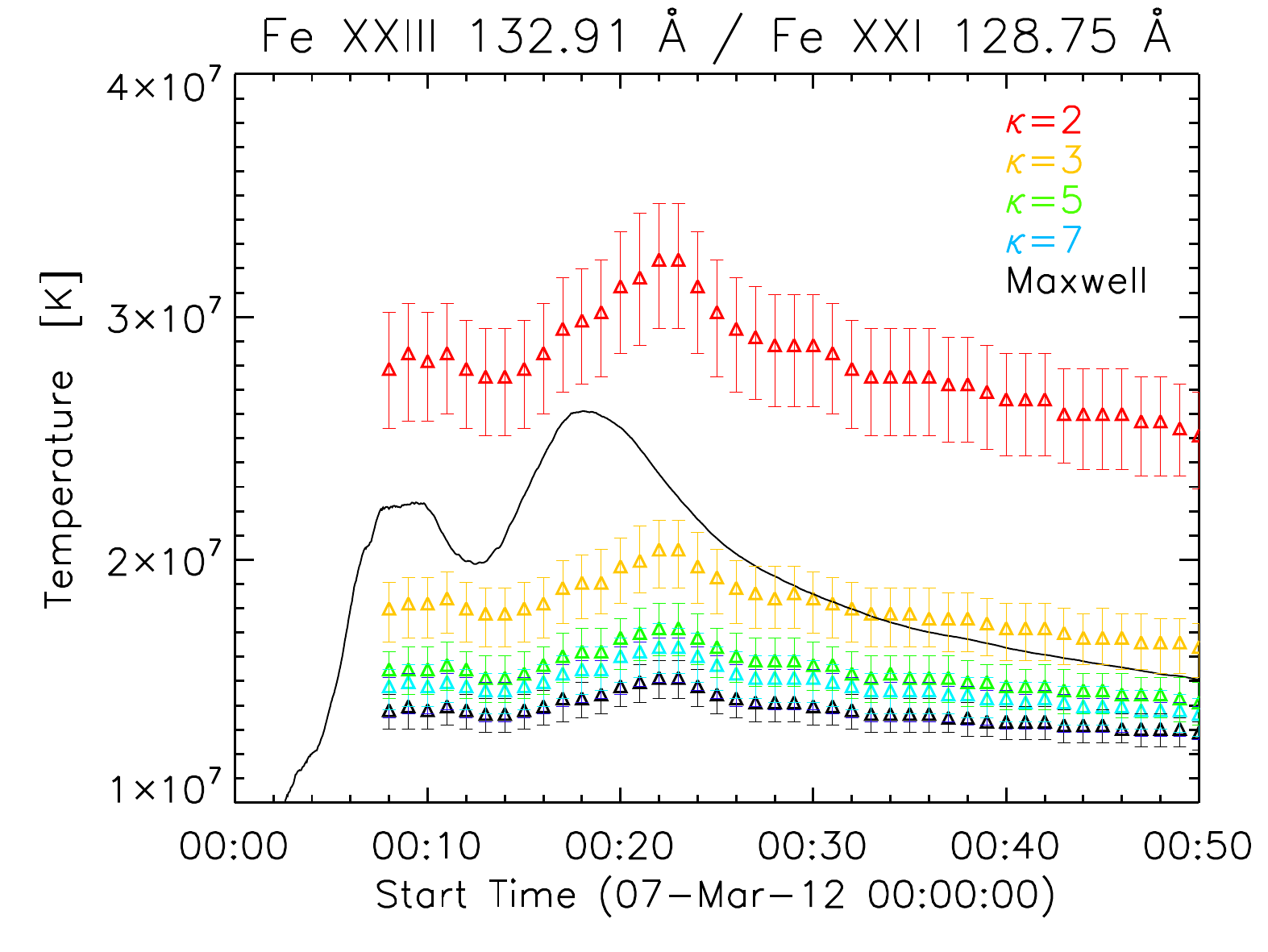}
	\includegraphics[width=7.5cm,clip]{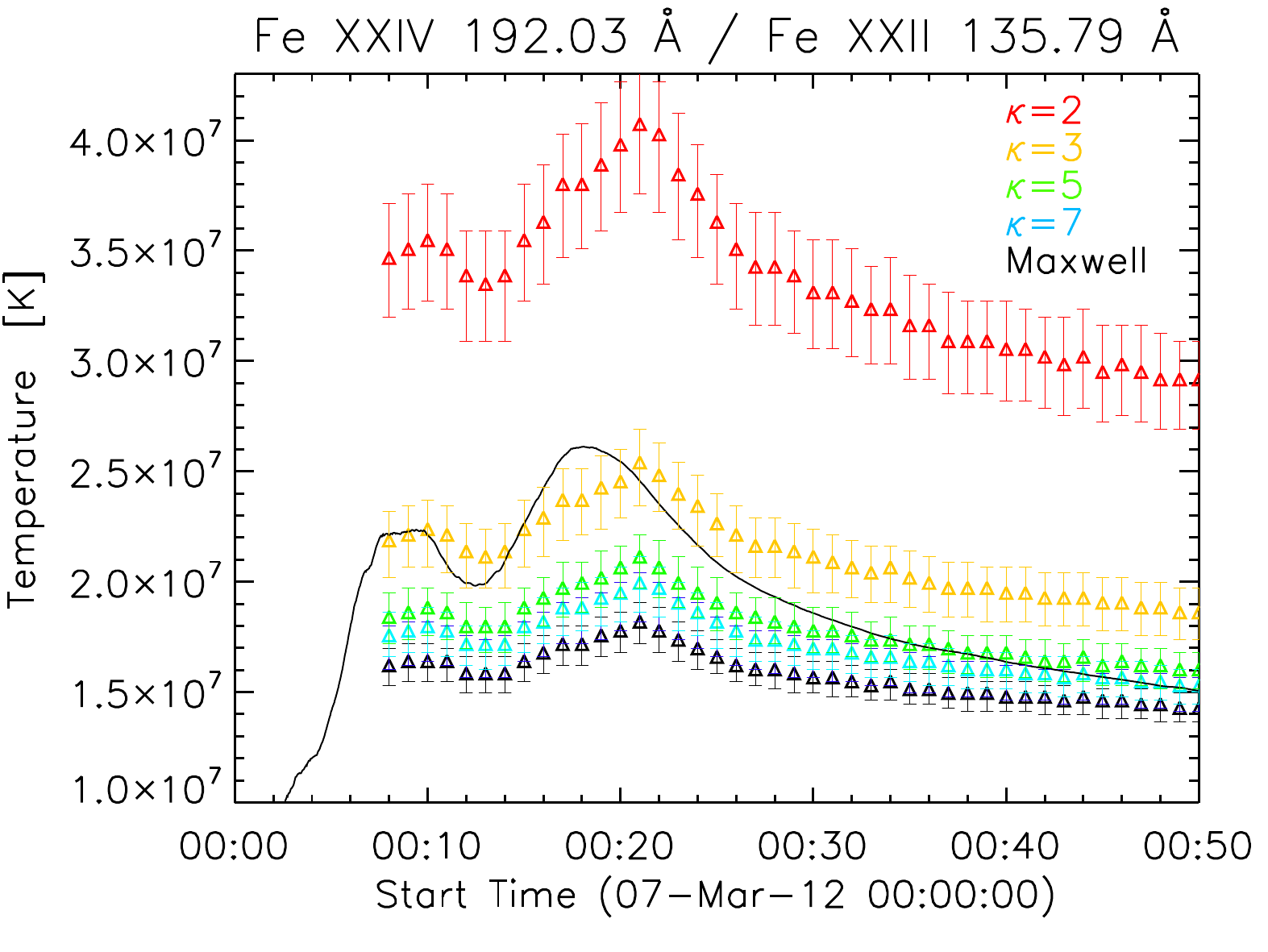}
\caption{Temperatures and their evolution diagnosed using the individual theoretical ratios from Fig. \ref{Fig:Diag_T_theor}. Colors denote the value of $\kappa$ as indicated. The thin black lines stand for the temperature derived from the ratio of the two GOES channels.}
\label{Fig:Diag_T}
\end{figure}
%
%
\section{Plasma diagnostics}
\label{Sect:4}

We now proceed to diagnose the basic plasma parameters: Electron density $n_\mathrm{e}$, electron temperature $T$, the $\kappa$ index, as well as the differential emission measure. To do this, line intensity ratios are used for diagnostics of $n_\mathrm{e}$, $T$, and $\kappa$, while lines spanning many ionization stages are used to diagnose the flare DEM$_\kappa(T)$.

Because of the low spectral resolution of EVE \citep[$\approx$1\,\AA,][]{Woods12} and the typical FWHM of flare lines is about 0.75\,\AA, most of the lines are blended \citep[][]{DelZanna13c}. Known blends from Fe flare lines were added to theoretical intensity calculations. To estimate the contribution of non-Fe unresolvable blends, we re-calculated the intensities of lines of interest including \textit{all} contributions included in the CHIANTI v7.1 and KAPPA databases. These contributions were calculated as a function of temperature and then folded over the DEM$_\kappa(T)$ derived from the flare (Sect. \ref{Sect:4.4}). These re-calculated contribution functions involving non-Fe were however not used for diagnostics, because of possible difficulties with anomalous abundances during flares \citep[see][]{Doschek15,Doschek16}. Subsequently, lines that were strongly blended by non-Fe lines were excluded from diagnostics. Weaker significant blends, as well as all other details on individual lines used for different types of diagnostics are discussed in Appendix \ref{Appendix:A}.

To enhance the signal-to-noise ratio in weaker lines, we performed averaging over 1 minute intervals. Furthermore, from each 1-minute averaged spectrum, we subtracted the pre-flare spectrum, which was obtained as an average over 3 minutes during the pre-flare period at 21:46 -- 21:49 UT, i.e., when the GOES X-ray signal was low, unperturbed by other flaring activity. These times are noted by blue dashed lines in Fig. \ref{Fig:Context}a. We note that subtracting the pre-flare spectrum is a standard practice for analysis of EVE spectra \citep[e.g.,][]{Milligan12,DelZanna13c}. This greatly helps to remove the blends from coronal and low-temperature lines. We also note that the coronal lines do not change strongly ($\lessapprox$20\%) during the flare, see Fig. 1 in \citet{DelZanna13c}.

Each subtracted spectrum was fitted using the XCFIT procedure available within SolarSoft, assuming a constant pseudo-continuum and Gaussian functions for each visible line feature including resolved blends within line wings. Details on the fitting of each line are also given in Appendix \ref{Appendix:A}. Finally, the uncertainties of the measured intensities include the photon noise uncertainty added in quadrature with the EVE calibration uncertainty, which is about 20\% \citep{Woods12}. For the weakest lines used here, the overall uncertainty can reach $\approx$40\%.

%
\subsection{Electron density diagnostics}
\label{Sect:4.1}


The electron density $n_\mathrm{e}$ is not a parameter of the distribution in Eq. (\ref{Eq:Kappa}) in the same manner as $T$ or $\kappa$. Thus, it can and should be diagnosed prior and separately from $T$ and $\kappa$ \citep{Dzifcakova10,Dudik14b,Dudik15}. Here, we use the well-known density-sensitive line ratio of \ion{Fe}{21} 145.73\,\AA\,/\,128.75\,\AA~\citep{Mason79,Mason84,Milligan12,DelZanna13c} which is density-sensitive above $\approx$10$^{11}$ cm$^{-3}$, i.e., in conditions corresponding to large flares \citep{Milligan12}. The sensitivity to $n_\mathrm{e}$ arises due to the presence of metastable levels within \ion{Fe}{21}, whose population is not strongly sensitive to either $T$ or $\kappa$. 

The theoretical line intensity ratio is shown in Fig. \ref{Fig:Diag_ne} \textit{top} as a function of $n_\mathrm{e}$. There, black and red colors denote Maxwellian and $\kappa$\,=\,2 distributions, respectively, i.e., the extreme values of the parameter $\kappa$ considered here. Individual linestyles denote different temperatures. The full lines correspond to the peak of the ionization equilibrium for a given $\kappa$, while the dashed and dash-dotted lines correspond to ion abundance being 10$^{-2}$ of the ion abundance peak. Thus, they denote the temperature interval where the ion is dominantly formed (see Fig. \ref{Fig:Ioneq}). Further quantification of the dependence of the \ion{Fe}{21} 145.73\,\AA\,/\,128.75\,\AA~ratio on $T$ and $\kappa$ is provided in Fig. \ref{Fig:Diag_ne} \textit{middle}. It is obvious that the line intensity ratio is not strongly sensitive to either $T$ or $\kappa$. 

Both \ion{Fe}{21} 128.75\,\AA~and 145.73\,\AA~lines are well observed by EVE \citep{Milligan12,DelZanna13c}. However, due to the relatively large uncertainty of the line intensities, the diagnosed electron densities (Fig. \ref{Fig:Diag_ne}, \textit{bottom}) have an uncertainty of about 0.3 in \logne. The densities diagnosed under the assumption of a Maxwellian or a $\kappa$-distribution are similar, about \logne\,$\approx$\,11.5; with the densities diagnosed for $\kappa$\,=\,2 being $\approx$0.15 dex smaller than those for the Maxwellian distribution. We note that this is a typical feature of the non-Maxwellian density diagnostics \citep[see][]{Dzifcakova10,Dudik14b,Dudik15}.

The observed densities do not evolve significantly during the flare, apart from perhaps a modest decrease with time. This decrease is however much smaller than the uncertainties of the diagnosed densities. We note that the observed density and its evolution is an average over the flare, since EVE is a full-Sun spectrometer. The absence of significant evolution is thus likely due to appearance of newer and newer flare loops. We also note that hydrodynamic models reflecting the overall evolution of many flare loops have been constructed, yielding similar absence of strong density evolution lasting for thousands of seconds \citep[e.g.,][]{Sun13,Polito16a}. In this respect, the reported density evolution is not unusual as it likely reflects a continuing energy release in the flare.

%
\subsection{Electron temperature diagnostics}
\label{Sect:4.2}

As mentioned in Sect. \ref{Sect:3}, both $T$ and $\kappa$ are free parameters of the distribution. Thus, diagnostics of temperature has to be done either simultaneously with $\kappa$ (Sect. \ref{Sect:4.3}), or under an \textit{assumption} of a constant $\kappa$. Since the latter is instructive in terms of influence of the $\kappa$-distributions on the observed spectra, we discuss this method first.

Under an assumption of constant $\kappa$, the diagnosed temperatures will necessarily depend on the assumed value of $\kappa$. This behavior comes primarily from the dependence of the ionization equilibrium on $\kappa$ (Fig. \ref{Fig:Ioneq}): The peaks of the relative ion abundance are wider and shifted to higher $T$ for lower $\kappa$ \citep{Dzifcakova13a}. Thus, the temperatures diagnosed for smaller $\kappa$ will be higher. 

To perform this diagnostics of $T$, we use the same three line intensity ratios as \citet{DelZanna13c}. These line ratios involve a pair of ions from different ionization stages, where the difference in ion charge is 2. This offers large sensitivity to $T$. The lines used are \ion{Fe}{20}\,121.85\,\AA, \ion{Fe}{21}\,128.75\,\AA, \ion{Fe}{22}\,135.79\,\AA, \ion{Fe}{23}\,132.91\,\AA, and \ion{Fe}{24}\,192.03\,\AA~(see Appendix \ref{Appendix:A} for details). The theoretical temperature-diagnostic curves for different $\kappa$ are shown in Fig. \ref{Fig:Diag_T_theor}. With progressively smaller $\kappa$, the curves are shifted to larger $T$ and are less steep, as expected.

The temperatures diagnosed using an assumption of constant $\kappa$ and the observed line ratios of \ion{Fe}{22} 135.79\,\AA\,/\,\ion{Fe}{20} 121.85\,\AA, \ion{Fe}{23} 132.91\,\AA\,/\,\ion{Fe}{21} 128.75\,\AA, and \ion{Fe}{24} 192.03\,\AA\,/\,\ion{Fe}{22} 135.79\,\AA~are shown in Fig. \ref{Fig:Diag_T} together with their respective uncertainties. The diagnosed temperatures indeed depend on the assumed value of $\kappa$. Progressively larger $T$ are obtained for smaller $\kappa$, with the $T_{\kappa=2}$ being about a factor of two higher than the $T_\mathrm{Maxw}$. This illustrates the importance of non-Maxwellian effects for diagnostics of temperature.

For comparison, the temperatures derived from the ratio of the two GOES channels (assuming Maxwellian) are shown as thin solid line. In accordance with the results of \citet{DelZanna13c}, it is seen that the GOES temperatures are discrepant from the Maxwellian temperatures by a factor of up to about two. That the $\kappa$-distributions yield higher temperatures for small $\kappa$ could hint at a possible resolution of this discrepancy. We however note that the $T_\mathrm{GOES}$ are likely a function of $\kappa$ as well, since at least the slope of the free-free continuum within the GOES pass-bands is a function of $\kappa$ at flare temperatures \citep[see Fig. 2 in][]{Dudik12}. The calculations of the GOES responses to $\kappa$-distributions, including contributions from various lines and continuum, is however out of the scope of the present work.

Except the dependence on $\kappa$, different line ratios yield different temperatures. In particular, using lines from ions from higher charge states yields higher temperatures. This likely reflects the fact that the EVE spectra are full-Sun and thus multithermal (see Sect. \ref{Sect:4.4}). We however note that the $\approx$14\% difference in the 192.03\,\AA~line intensities between CHIANTI v7.1 and v8 is approximately sufficient to bring the temperatures from the \ion{Fe}{24}\,/\,\ion{Fe}{22} ratio into agreement with the \ion{Fe}{23}\,/\,\ion{Fe}{21} one.

The temperatures show a clear evolution during the flare (Fig. \ref{Fig:Diag_T}). Initial temperatures diagnosed from the line ratios start above 10\,MK for Maxwellian and above 20\,MK for $\kappa$\,=\,2, respectively. A rise is detected at about 00:12\,UT, lasting until the peak at about 00:22\,UT. This peak occurs after the strongest gradient of the X-ray flux during the impulsive phase (at about 00:18\,UT, see Fig. \ref{Fig:Context}), but before the peak of the X-ray flux at 00:25\,UT. After 00:22\,UT, the temperature decreases steadily, with the temperatures diagnosed from \ion{Fe}{24}\,/\,\ion{Fe}{22} dropping faster those diagnosed from other ratios involving lower charge states.

%
%
\subsection{Diagnostics of the non-Maxwellian parameter $\kappa$}
\label{Sect:4.3}

The parameter $\kappa$ has to be diagnosed simultaneously with $T$, since both are parameters of the distribution (see Eq. \ref{Eq:Kappa}). To do this, we use the ratio-ratio method \citep{Dzifcakova10,Dudik14b,Dudik15}, where a dependence of one line ratio is plotted against a different line ratio. Typically, one ratio is chosen to include single-ion lines either separated in wavelength or with different behavior of the excitation cross-section with $E$. Such ratios are sensitive to $\kappa$ since the two lines are excited by different parts of the distribution. The second ratio typically involves temperature-sensitive lines. Using lines from two neighboring ionization stages increases the sensitivity to $T$ but could introduce uncertainties due to the non-equilibrium ionization. However, we note that the high flare densities mean that the plasma is expected to be close to ionization equilibrium, especially after the early phase of the flare, see \citet{Smith10} and \citet[][and references therein]{Dudik17a}.

\begin{figure*}[!ht]
	\centering
	\includegraphics[width=8.8cm,clip]{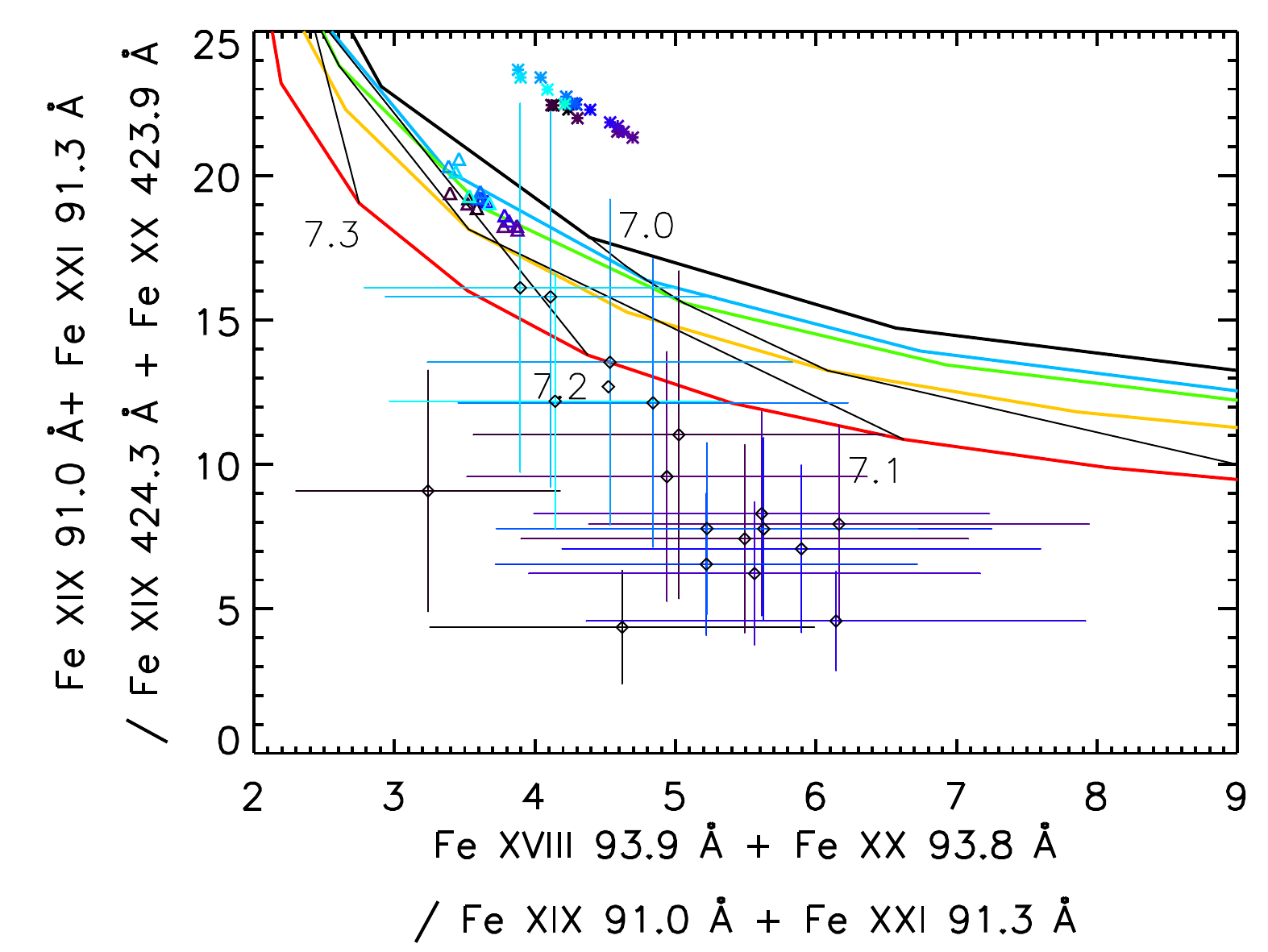}
	\includegraphics[width=8.8cm,clip]{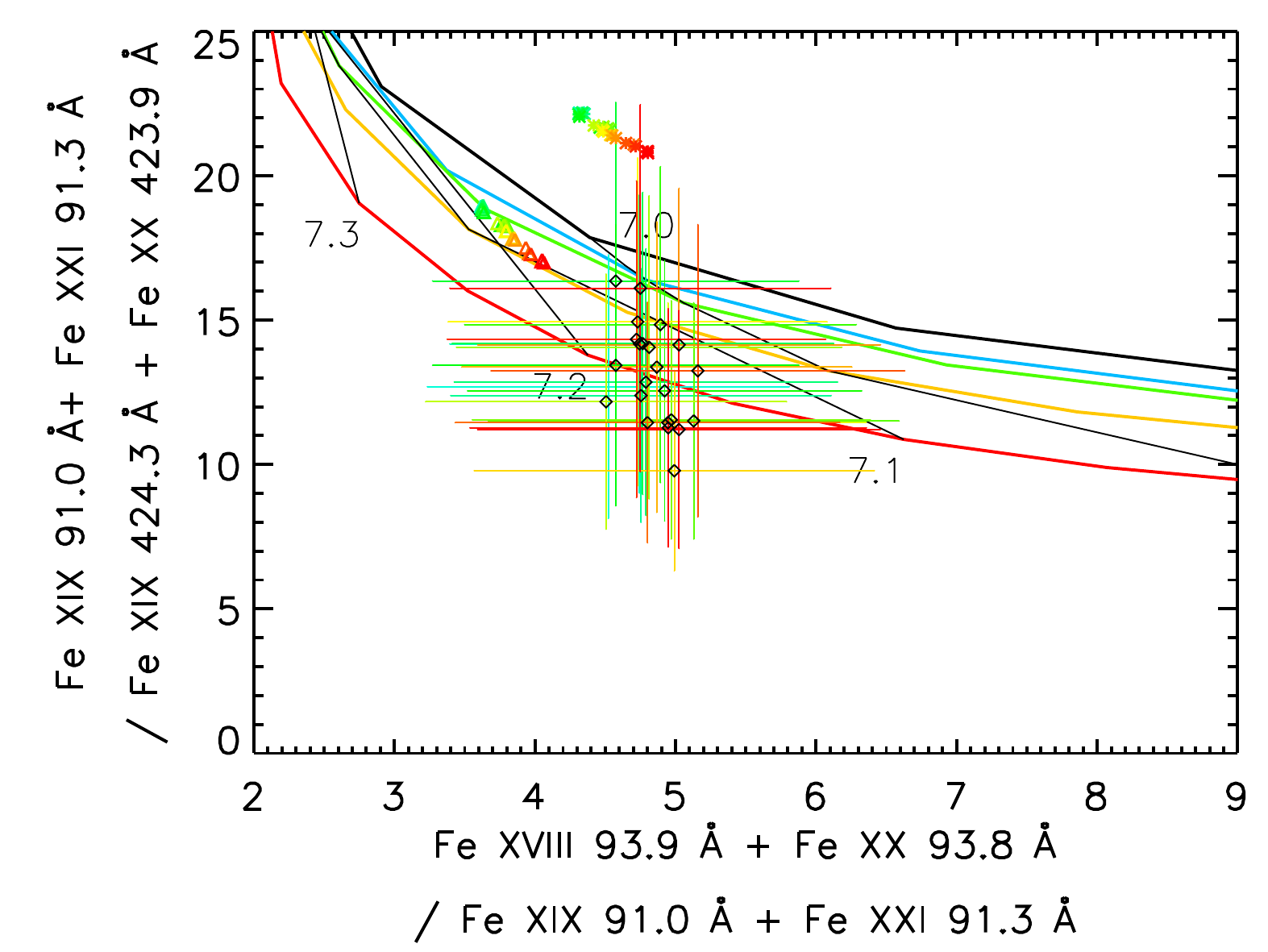}
	\includegraphics[width=17.6cm,clip]{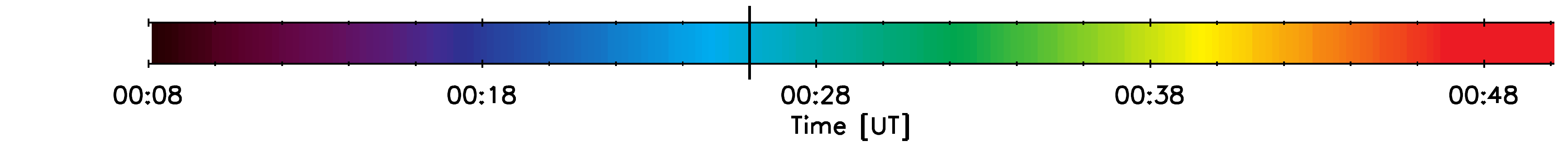}
	\includegraphics[width=8.8cm,clip]{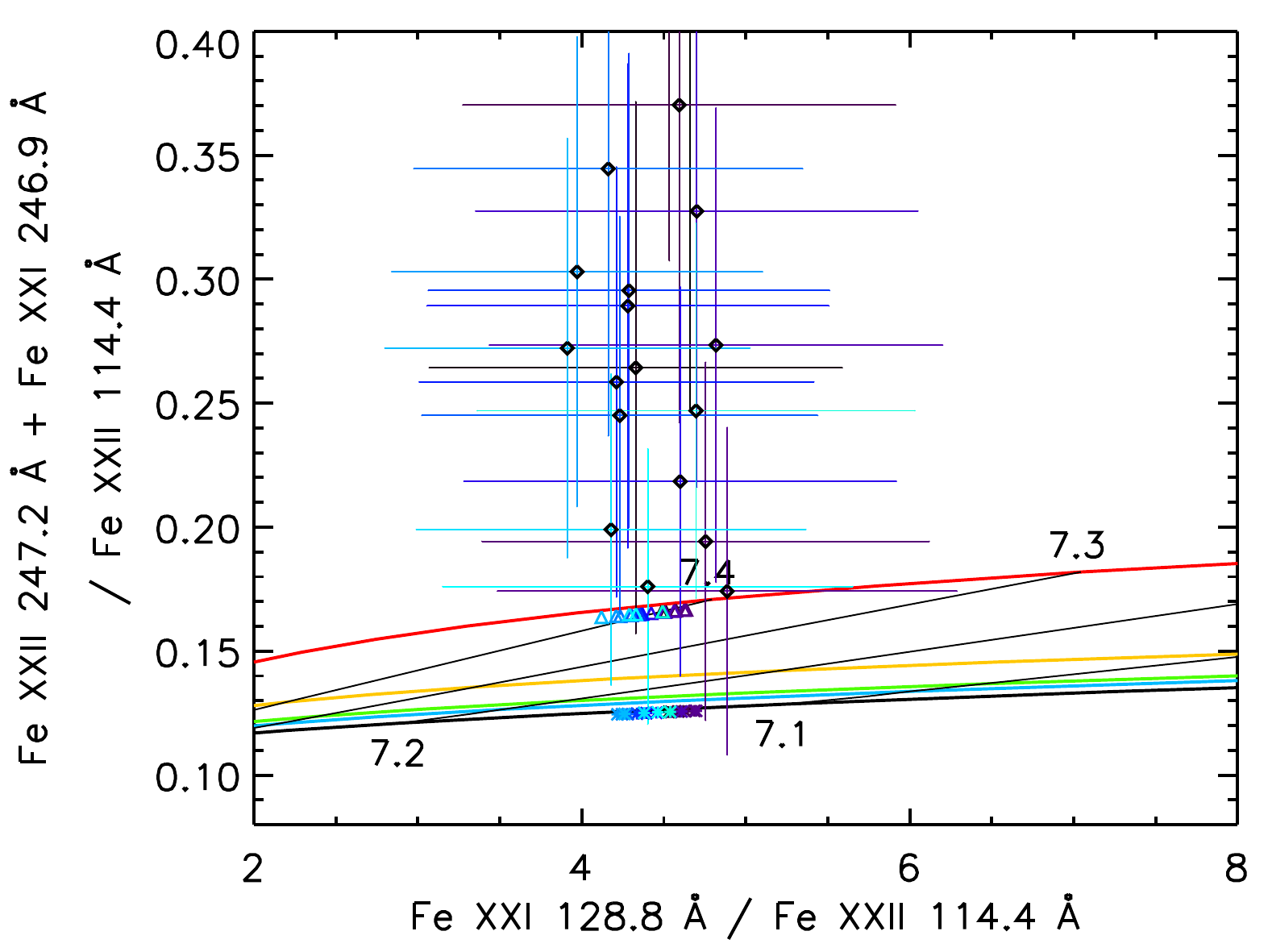}
	\includegraphics[width=8.8cm,clip]{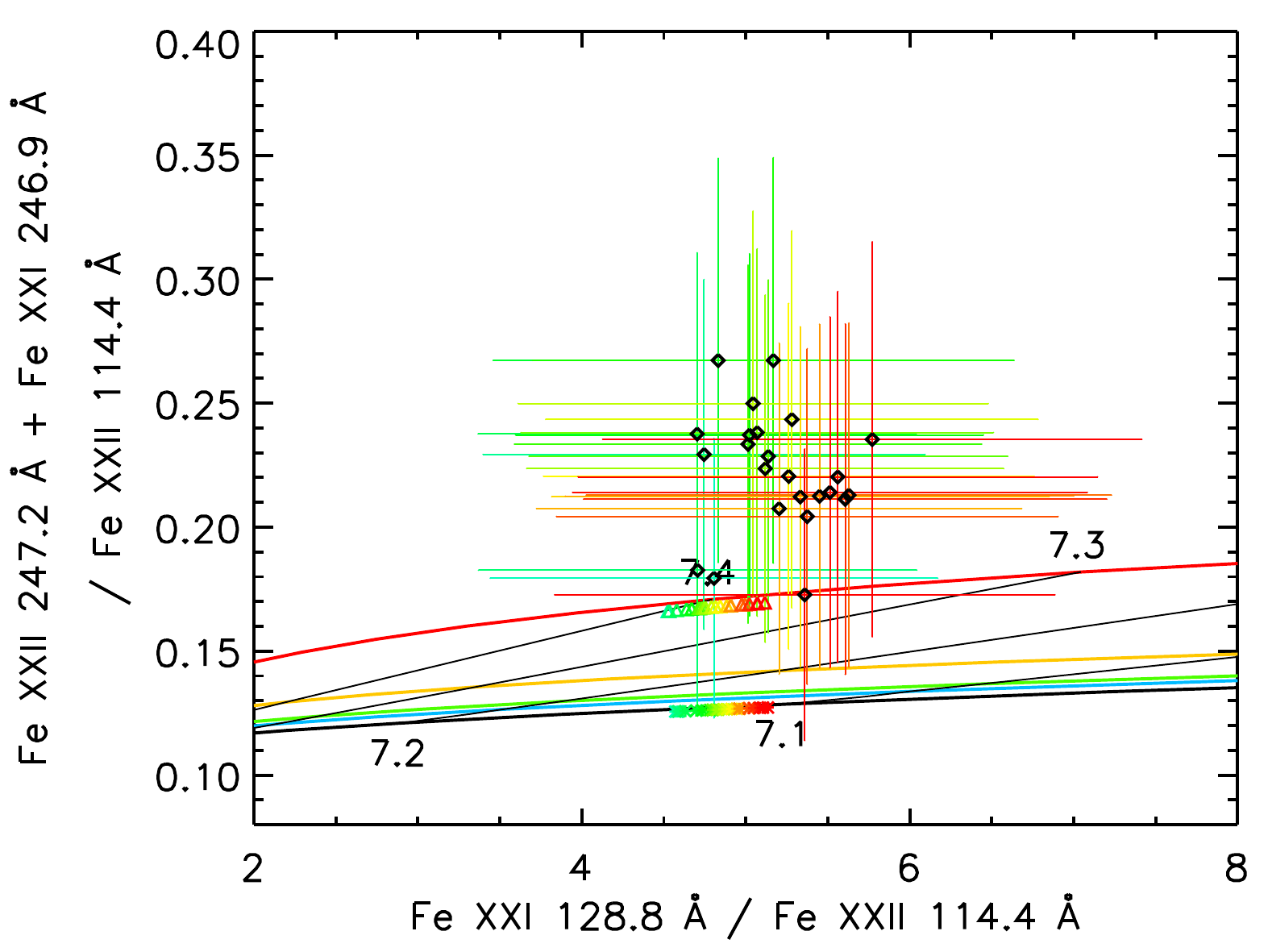}
\caption{Diagnostics of the electron distribution. Colored curves give the theoretical ratios as a function of $\kappa$. Maxwellian is black, while $\kappa$\,=\,2 is red. Thin black lines denote isotherms, with the value of \logt~indicated. The crosses indicate the observed line intensity ratios and their uncertainties. The color of the crosses denote time during the flare (00:08 -- 00:50\,UT), as indicated by the color bar. The \textit{left} panels show results until the peak at 00:25\,UT, while the \textit{right} images show the results during 00:26--00:50\,UT. Colored asterisks and triangles denote the theoretical line ratios for Maxwellian and $\kappa$\,=\,2, folded over the respective DEM$_\kappa(T)$.}
\label{Fig:Diag_kappa}
\end{figure*}

To diagnose $\kappa$ and $T$ simultaneously, we use a ratio-ratio diagram based on lines of \ion{Fe}{19} combined with \ion{Fe}{18}, and an additional ratio-ratio diagram using \ion{Fe}{22} in combination with \ion{Fe}{21}. These ratio-ratio diagrams are shown in Fig. \ref{Fig:Diag_kappa}.

%

%
\subsubsection{Diagnostics from \ion{Fe}{19}}
\label{Sect:4.3.1}

The sensitivity to $\kappa$ in the first ratio-ratio diagram arises from a combination of the \ion{Fe}{19} lines at 91.0\,\AA~and 424.3\,\AA, which are widely separated in wavelength. Both these lines are blended with either \ion{Fe}{20} or \ion{Fe}{21}. For details, see Appendix \ref{Appendix:A_19}. The sensitivity to temperature is obtained from a combination of the 91.0\,\AA~blend with the well-known \ion{Fe}{18} line at 93.9\,\AA, which is also blended with \ion{Fe}{20} (Appendix \ref{Appendix:A_18}). The theoretical ratio-ratio diagram is shown in Fig. \ref{Fig:Diag_kappa}, \textit{top}. There, the curves shown by full lines denote the theoretical ratios as a function of $\kappa$, with their color-coding is the same as in Figs. \ref{Fig:Diag_T_theor} and \ref{Fig:Diag_T}. In particular, black lines represent Maxwellian theoretical ratios, and red represents $\kappa$\,=\,2. Isotherms connecting points with different $\kappa$ but the same \logt~are indicated as thin black lines.

The observed ratios and their 1--$\sigma$ uncertainties are shown by diamonds and colored crosses, where the color indicates the time during the flare, starting from 00:08\,UT (black and violet) to 00:50\,UT (red). Diagnostics prior to 00:80\,UT is not possible since the weaker lines cannot be identified in the observed spectra before this time. After 00:80\,UT, the results of the diagnostics indicate a range of $\kappa$ values, depending on the time. We note that accurate determination of a $\kappa$ value is not possible due to the large observational uncertainties with respect to the spread of the theoretical curves for different $\kappa$ values. Within the limit of the uncertainties, we can only determine that the plasma is likely extremely non-Maxwellian, with values of $\kappa$\,$\lesssim$\,2 diagnosed during the early and impulsive phases of the flare, from 00:08\,UT to approximately 00:20\,UT (black, violet, and blue; Fig. \ref{Fig:Diag_kappa}, \textit{top left}). Subsequently, the plasma thermalizes, with the yellow to red crosses being consistent even with the Maxwellian distribution within their respective uncertainties.

We further note that at around 00:80\,UT, some of the observed ratios (black crosses) are far from the diagnostic curves. The cause of this is not clear. It could indicate possible departures of the true electron energy distribution from a $\kappa$-distribution at the start of the flare, or problems with blends or identifications of weaker lines.

Finally, we note that the $T$ diagnosed simultaneously with $\kappa$ is lower than those obtained in Sect. \ref{Sect:4.2}. In our case, we obtain \logt\,$\approx$\,7.1--7.2, which is likely caused by the plasma multithermality: The values of $T$ obtained from the ratio-ratio diagrams reflect the $G_{X,ji}(T,n_\mathrm{e},\kappa)$ of the ions used to diagnose $T$ and $\kappa$ simultaneously.

\subsubsection{Diagnostics from \ion{Fe}{22}}
\label{Sect:4.3.2}

To verify the results of the non-Maxwellian diagnostics, as well as to perform it from lines formed at higher $T$, we use the \ion{Fe}{22} ratio-ratio diagrams. The ratio \ion{Fe}{22} 247.2\,\AA\,/\,114.4\,\AA~is sensitive to $\kappa$ since it involves lines formed at wavelengths different by about a factor of two. The \ion{Fe}{22} 247.2\,\AA~line is however blended with \ion{Fe}{21} (Appendix \ref{Appendix:A_22}). The sensitivity to $T$ comes from the combination of \ion{Fe}{22} 114.4\,\AA~with \ion{Fe}{21} 128.75\,\AA. 

The theoretical ratio-ratio curves together with the observed intensities are shown in Fig. \ref{Fig:Diag_kappa}, \textit{bottom}. Overall, the results confirm the picture obtained from \ion{Fe}{19}. It is again seen that the plasma is strongly non-Maxwellian during the early and impulsive phases of the flare, while the plasma becomes closer to Maxwellian during the peak and gradual phases. The observed points are however further away from the theoretical ratios than in the case of diagnostics from \ion{Fe}{19}. This could be at least in part due to the unresolved AR blend of \ion{S}{11} at 246.90\,\AA~(Appendix \ref{Appendix:A_22}) that was not included in the theoretical intensity calculations of the \ion{Fe}{22}+\ion{Fe}{21} blend at 247.2\,\AA. This is since sulfur is not a low-FIP element as iron, and thus it could possibly experience anomalous abundances during flares \citep[c.f.,][]{Doschek15,Doschek16}.

%
\begin{figure}
	\centering
	\includegraphics[width=8.8cm,clip]{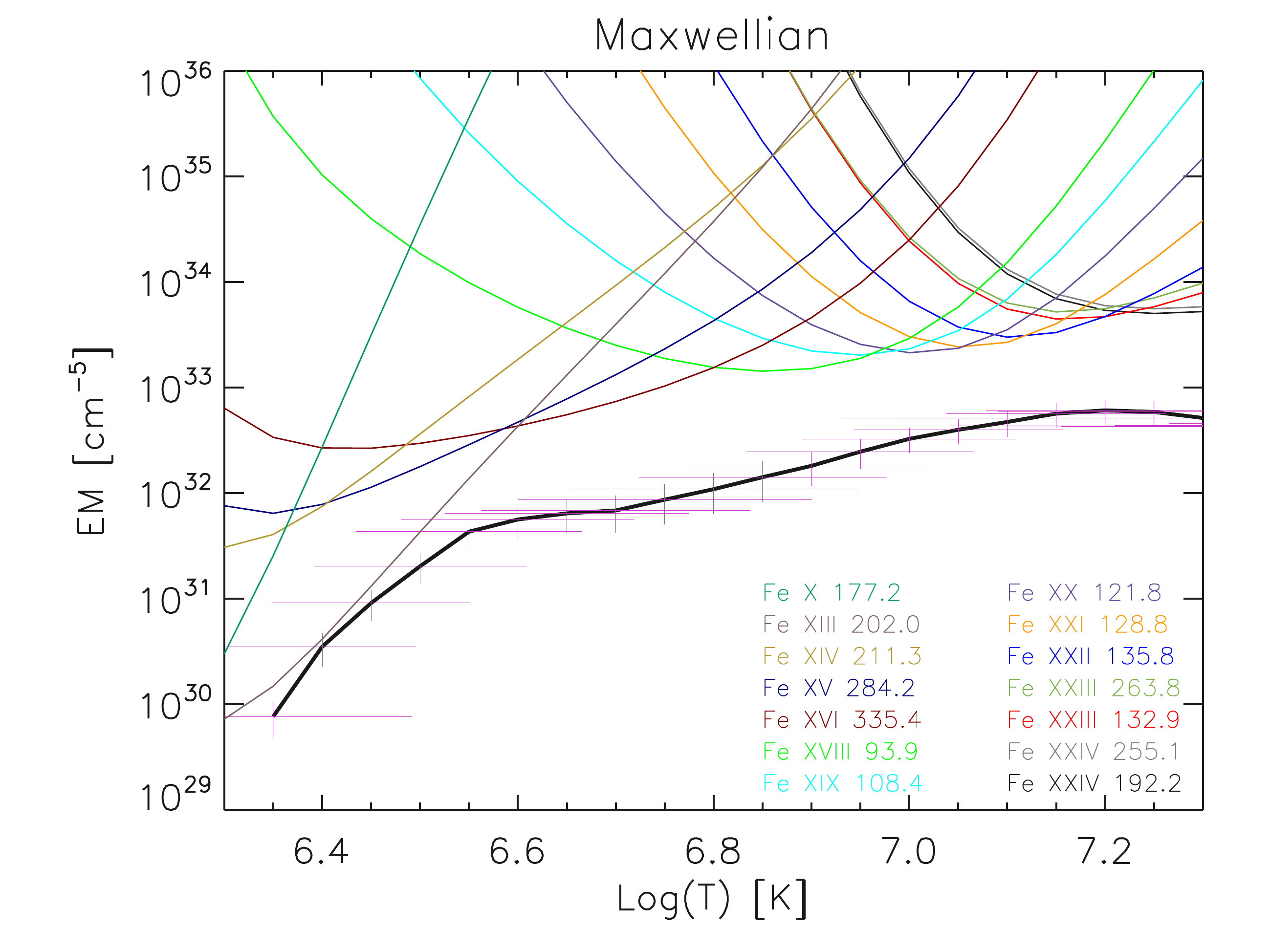}
	\includegraphics[width=8.8cm,clip]{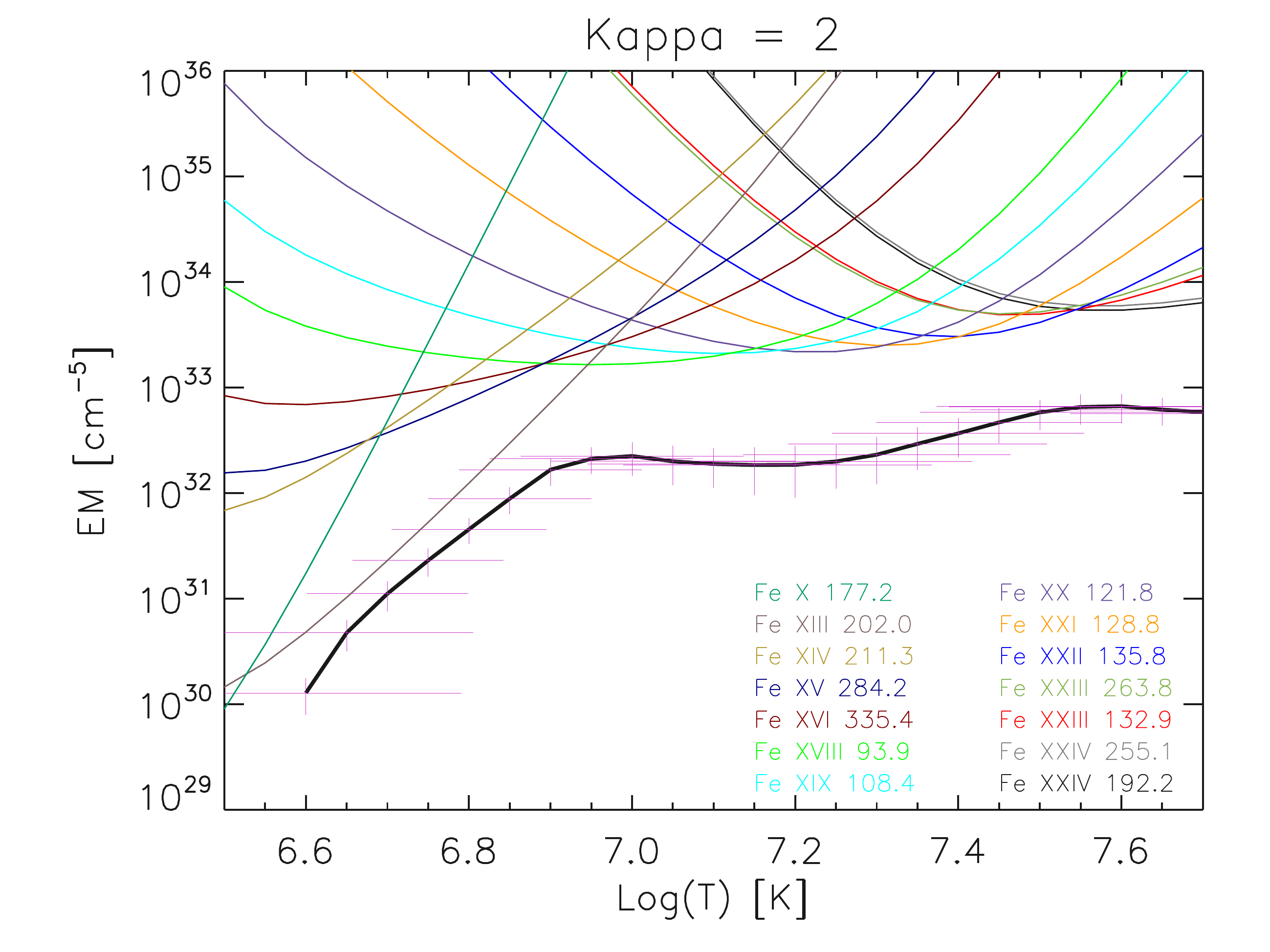}
\caption{Emission measure distributions and its uncertainties derived from the spectrum at 00:25\,UT together with the respective EM-loci plots. The spectral lines used are indicated. \textit{Top}: EM$_\mathrm{Maxw}(T)$ derived for the Maxwellian distribution. \textit{Bottom}: The $EM_{\kappa=2}(T)$ derived for $\kappa$\,=\,2.}
\label{Fig:DEM_25}
\end{figure}
%
\begin{figure}
	\centering
	\includegraphics[width=8.8cm,clip]{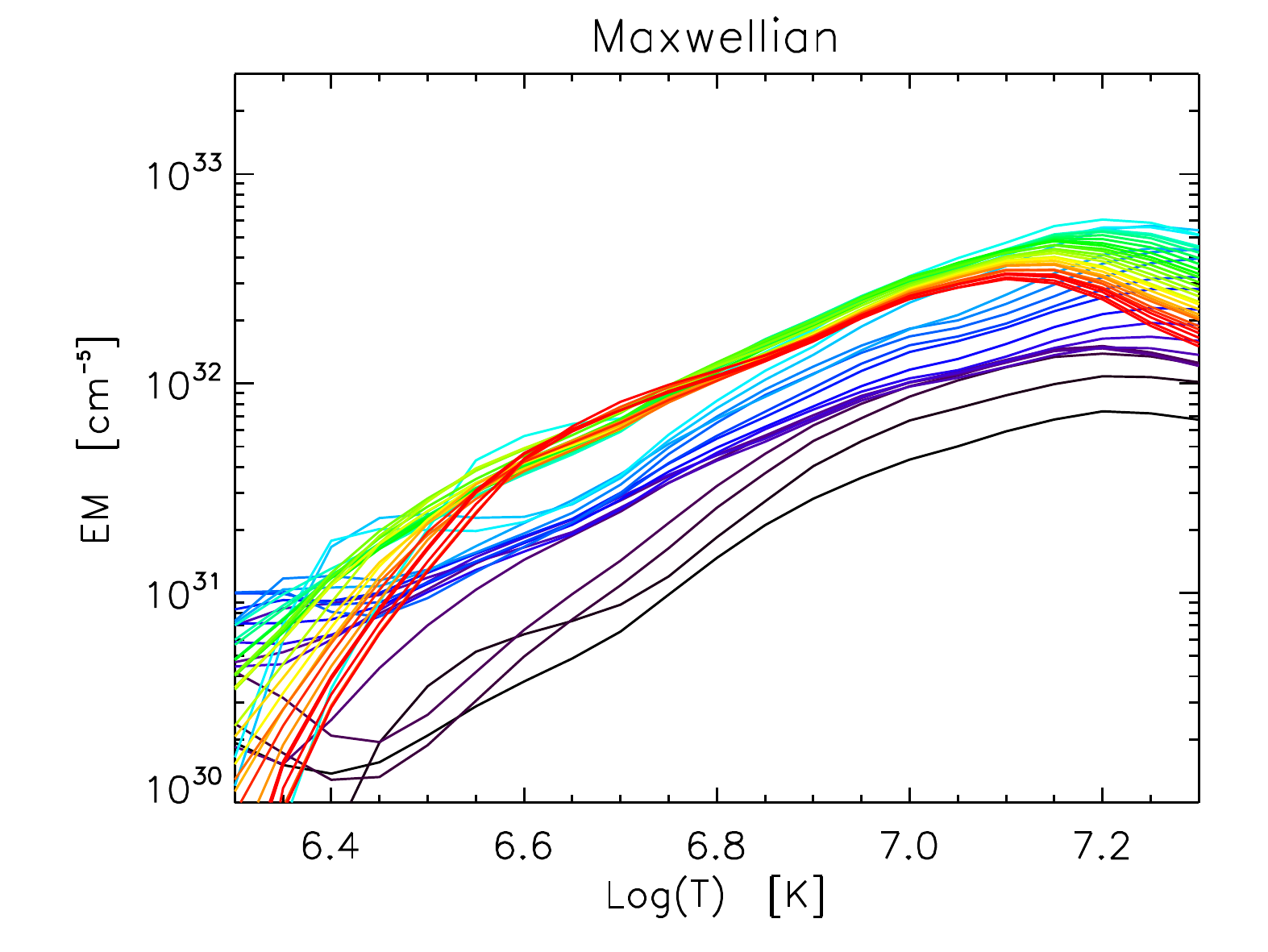}
	\includegraphics[width=8.8cm,clip]{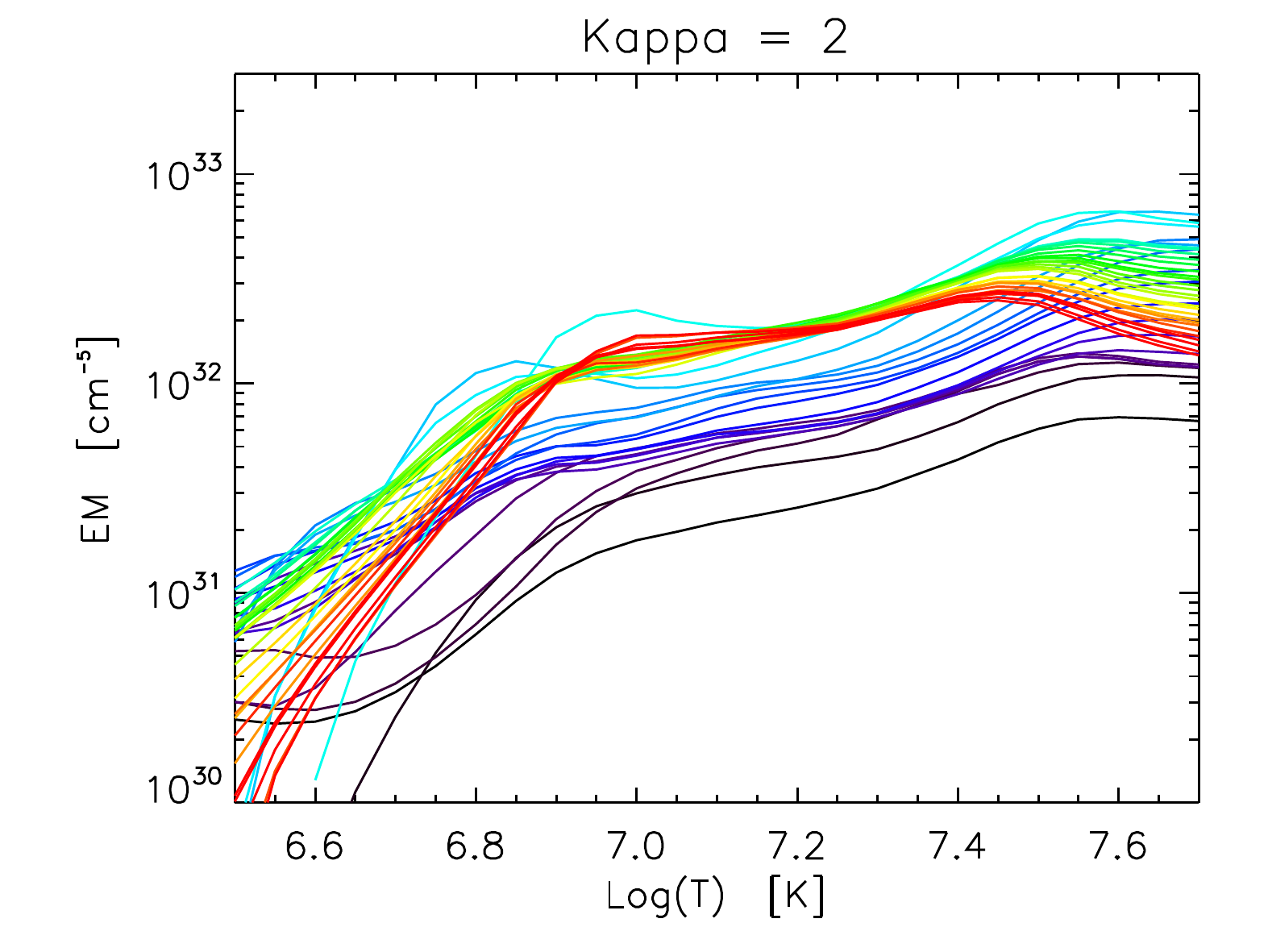}
	\includegraphics[width=8.8cm,clip]{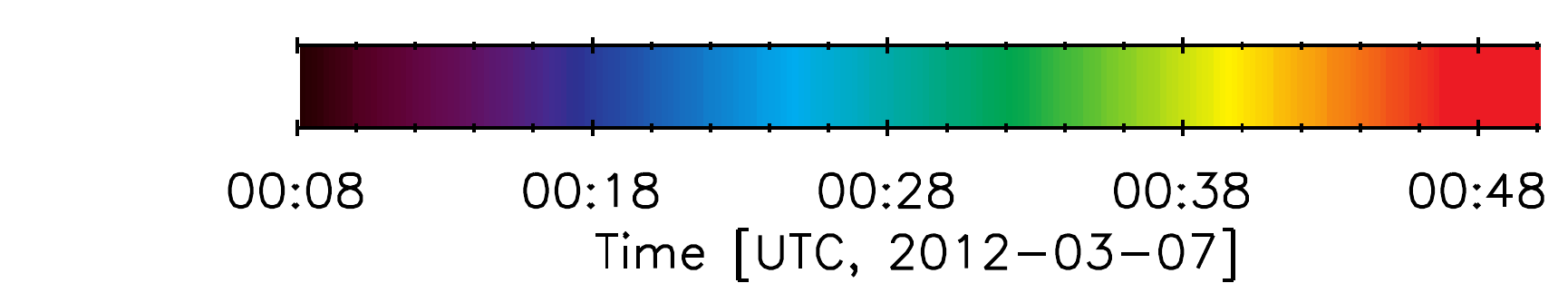}
\caption{Evolution of the EM(T) distribution during the studied time interval. The colors denotes time during the flare.}
\label{Fig:DEM_evol}
\end{figure}
%
%
\subsection{Differential emission measure}
\label{Sect:4.4}

In previous Sections \ref{Sect:4.2} and \ref{Sect:4.3}, we obtained different temperatures using line ratios from different ionization stages, suggesting that the plasma is multithermal. This is not surprising, since EVE is a full-Sun spectrometer, and the flare is an eruptive one, i.e., it involves multiple emitting flare loops (Fig. \ref{Fig:Context}c--e).

To quantify the degree of multithermality of the flare, we performed DEM$_\kappa(T)$ analysis using the regularization inversion method of \citet{Hannah12}. This method can be straightforwardly generalized for non-Maxwellian distributions simply by supplying it with the non-Maxwellian $G_{X,ji}(T,n_\mathrm{e},\kappa)$ \citep{Mackovjak14}. The advantage of this method is that it provides not only the DEM$_\kappa(T)$, but also its uncertainties in both the DEM and $T$ as well. To reconstruct the DEM, we used the EVE flare lines together with additional well-known lower-$T$ EUV lines to constrain the temperature space. These lower-$T$ lines include the \ion{Fe}{10} 177.2\,\AA, \ion{Fe}{13} 202.0\,\AA, \ion{Fe}{14} 211.3\,\AA, \ion{Fe}{15} 284.2\,\AA, and \ion{Fe}{16} 335.4\,\AA~lines \citep[e.g.,][]{ODwyer10,Warren12,DelZanna13b}. Especially the \ion{Fe}{10} and \ion{Fe}{13} provide strong constraints (Fig. \ref{Fig:DEM_25}) for temperatures below \logt\,=\,6.3 and 6.6 for Maxwellian and $\kappa$\,=\,2, respectively. In addition, the line of \ion{Fe}{19} 108.4\,\AA~was used instead of other \ion{Fe}{19} lines, since this is the best line for EM analyses, see \citet{DelZanna13c}. 

We first applied this method to obtain the DEM$_\kappa(T)$ for the flare peak at 00:25\,UT. The corresponding emission measure distributions EM$_\kappa(T)$, obtained as DEM$_\kappa(T)\mathrm{d}T$, are shown in Fig. \ref{Fig:DEM_25}. There, the EM$_\kappa(T)$ are shown for the Maxwellian and $\kappa$\,=\,2 together with the respective EM-loci plots \citep[][]{Strong78,Veck84,DelZanna03,Mackovjak14}. The EM$_\kappa(T)$ obtained for these two distributions are similar, except a shift towards higher $T$ for $\kappa$\,=\,2, which occurs mainly as a result of the behavior of the ionization equilibrium with $\kappa$ (see Fig. \ref{Fig:Ioneq}).

Both EM$_\kappa(T)$ distributions are relatively flat at temperatures above \logt\,=\,6.6 and 6.9 for the Maxwellian and $\kappa$\,=\,2, respectively, indicating strongly multithermal plasma. Their peak occurs at about \logt\,=\,7.2 and 7.5, respectively, and the EM$_\kappa(T)$ decrease only about an order of magnitude between \logt\,=\,6.6--7.2 (Maxwellian) and 6.9--7.6 ($\kappa$\,=\,2). At lower $T$, the EM$_\kappa(T)$ decrease sharply, mainly as a result of the subtraction of the pre-flare spectrum. It is this decrease that suppresses the QS and AR blends to EVE flare lines (see Appendix \ref{Appendix:A} for details).

We note that the high-temperature end of the DEM$_\kappa(T)$ is poorly constrained due to lack of EVE lines formed at temperatures above the formation temperature of \ion{Fe}{24}. This ion has maximum of the relative ionization abundance at \logt\,=\,7.25 for Maxwellian and 7.60 for $\kappa$\,=\,2. This increase with $\kappa$ occurs due to large increase in both ionization and recombination rates with decreasing $\kappa$. Subsequently, the peak of the DEM$_\kappa(T)$ found for this flare occurs close to peak formation temperature of \ion{Fe}{24}, i.e., \logt\,=\,7.25 for Maxwellian and 7.60 for $\kappa$\,=\,2 (Figs. \ref{Fig:DEM_25} and \ref{Fig:DEM_evol}). Such peaks are by necessity poorly constrained as well. However, without observations of the flare plasma in X-rays from either RHESSI or the X-ray Telescope \citep[XRT,][]{Golub07} onboard Hinode/XRT, it is not possible to provide accurate high-temperature constraints for our DEMs.

The evolution of the EM$_\kappa(T)$ during the flare is shown in Fig. \ref{Fig:DEM_evol}. There, the color-coding of the EM$_\kappa(T)$ curves is the same as in Fig. \ref{Fig:Diag_kappa}. It is seen that the flat EM$_\kappa(T)$ curves persist throughout the 00:08--00:50\,UT period, and that their character do not change strongly over time, except for an increase in the overall emission measure, by about an order of magnitude. Largest EMs occur at about 00:25\,UT; i.e., during the peak phase of the soft X-ray flux (c.f., Fig. \ref{Fig:Context}), and persist until the end of the analyzed period 25\,min later. Finally, the maxima of the EM$_\kappa(T)$ curves at \logt\,$\approx$\,7.2 for the Maxwellian and \logt\,$\approx$\,7.6 for $\kappa$\,=\,2 occur already at the start of the flare at about 00:08\,UT, and persist until about the peak phase at 00:25\,UT (blue and cyan color). These maxima are however poorly constrained as discussed above. At later times, the maxima decrease to \logt\,=\,7.1 and 7.4 during the gradual phase (red curves) for the Maxwellian and $\kappa$\,=\,2, respectively.

%
\subsection{Influence of DEM$_\kappa(T)$ on the diagnostics of $\kappa$}
\label{Sect:4.5}

Since the ratio-ratio diagrams used to diagnose $\kappa$ in Sect. \ref{Sect:4.3} were constructed under the isothermal assumption, we used the DEM$_\kappa(T)$ obtained in Sect. \ref{Sect:4.4} to calculate the DEM-predicted diagnostic line ratios as a function of $\kappa$ (see Eq. \ref{Eq:line_intensity_DEM}) and time. This is useful in estimating the theoretical ratios as a function of $\kappa$ for multithermal situations; and it is the same procedure as outlined in \citet[][Sect. 4.3.2 therein]{Dudik15}.

Effectively, the DEM-predicted line ratios are the ratio-ratio curves weighted by the DEM$_\kappa(T)$. I.e., the DEM-predicted ratios will always lie close to the theoretical curves in the ratio-ratio diagrams; departure from the curve is possible only in the local direction of curvature. This also means that for flat curves, the DEM-predicted ratios will lie very close to the respective curves.

This is indeed what we found. In Fig. \ref{Fig:Diag_kappa}, the DEM-predicted ratios are shown for the Maxwellian and $\kappa$\,=\,2 as colored asterisks (for Maxwellian) and triangles (for $\kappa$\,=\,2), where the color again stands for time. Since the DEM$_\kappa(T)$ do not change their shape appreciably, the DEM-predicted ratios for different times are clustered. The ratio-ratio diagram involving \ion{Fe}{19} lines contains diagnostic curves that are locally convex; the DEM-predicted ratios then lie above these curves. In particular, the DEM-predicted ratios for $\kappa$\,=\,2 lie at the ratio-ratio curves corresponding to $\kappa$\,=\,3--7. However, we note that the distance between the DEM-predicted ratios for the Maxwellian and $\kappa$\,=\,2 distribution is much smaller than the size of the error-bar of individual observed ratios. This means that only the observed ratios far from the DEM-predicted ratios can be confidently described as strongly non-Maxwellian. These are the ratios detected during the flare start and impulsive phases at about 00:08--00:20\,UT (Sect. \ref{Sect:4.3.1}). 

The ratio-ratio diagram involving \ion{Fe}{22} has much flatter ratio-ratio curves than the one involving \ion{Fe}{19}, and the DEM-predicted ratios for Maxwellian and $\kappa$\,=\,2 lie very close to the respective curves as expected. Again, this increases the confidence that the ratios during the start and impulsive phases of the flare are strongly non-Maxwellian.

We however note that some ratios observed at the very start of the flare (black crosses in Fig. \ref{Fig:Diag_kappa}) are very far from the respective DEM-predicted ones even for $\kappa$\,=\,2. In the case of \ion{Fe}{22}, they are away from the DEM-predicted ones by as much as a factor of 2.5. The reason for this is unknown. It is unlikely to be due to the known QS or AR blends (see, e.g., Appendix \ref{Appendix:A}). It is possible either that at the start of the flare, the true electron energy distribution is not well-described by a $\kappa$-distribution, or there are other effects at play, such as non-equilibrium ionization, or both. We note that the ionization equilibration timescales in flares can be of the order of seconds, tens of seconds, or possibly even minutes depending on the electron density \citep{Doschek79,Doschek87,Golub89,Bradshaw04,Smith10,Polito16a}.
The high densities above \logne\,=\,11 diagnosed in Sect. \ref{Sect:4.1} should strongly suppress the non-equilibrium ionization effects.

%
%
\section{Summary}
\label{Sect:5}

We performed the non-Maxwellian diagnostics of the eruptive X5.4-class flare of 2012 March 07 using full-Sun spectra observed by the SDO/EVE instrument. The spectra were averaged over 1 minute during 00:08--00:50\,UT and a pre-flare spectrum was subtracted. Theoretical line intensity calculations were performed for the non-Maxwellian $\kappa$-distributions by using the KAPPA database. While these distributions might not totally describe the evolving flare plasma, they allow for modeling of the effect of high-energy tails on the spectra at the expense of only one extra parameter, $\kappa$. The theoretical non-Maxwellian line intensity calculations were compared with the observed ones for a range of ions, from \ion{Fe}{19} to \ion{Fe}{24}. The main findings can be summarized as follows:
\begin{enumerate}
 \item The electron densities diagnosed using \ion{Fe}{21} reach \logne\,=\,11.5, and do not evolve strongly during the entire studied time interval. The \ion{Fe}{21} 145.7\,/\,128.8\,\AA~ratio is not strongly sensitive to either $T$ or $\kappa$, making the electron densities the only plasma parameter independent of the other ones.
 \item The temperatures diagnosed under an assumption of a constant $\kappa$ depend strongly on the assumed value of $\kappa$. This is a consequence mainly of the behavior of the ionization equilibrium. The temperatures diagnosed for $\kappa$\,=\,2 are about a factor of 2 higher than the Maxwellian temperatures.
 Additionally, the temperatures depend on the line ratios used, with \ion{Fe}{22}/\ion{Fe}{20}, \ion{Fe}{23}/\ion{Fe}{21} and \ion{Fe}{24}/\ion{Fe}{22} yielding progressively higher temperatures, which is a signature of multithermality.
 \item Maxwellian temperatures diagnosed from line ratios are inconsistent with the temperature derived from a ratio of GOES channels. The GOES temperatures are higher than those from the line ratios, as found already by \citet{DelZanna13c}. We suggest that the $\kappa$-distributions could represent a possible resolution of this discrepancy.
 \item The temperatures evolve during the flare, rising at 00:12\,UT and peaking at 00:22\,UT, after the strongest gradient of the X-ray flux during the impulsive phase, but before the peak of the soft X-ray flux as detected by GOES. The temperatures then decrease afterwards.
 \item Extremely non-Maxwellian values of $\kappa$\,$\lesssim$\,2 are diagnosed until about 00:20\,UT, i.e., during the early and impulsive phases of the flare. Subsequently, the plasma thermalizes, i.e., moves closer to Maxwellian. The error-bars of the observed ratios are however large compared to the spread of the curves for diagnostics of $\kappa$, which precludes determination of $\kappa$ after about 00:20\,UT, i.e., during the thermalization.
 \item The plasma is found to be multithermal, with relatively flat DEM$_\kappa(T)$ independently of the $\kappa$ value used. The shape of the DEM does not strongly evolve with time, except an overall increase of the total emission measure, and decrease of its peak. The peak occurs at \logt\,$\approx$\,7.2 and 7.6 for the Maxwellian and $\kappa$\,=\,2 during the early and impulsive phases of the flare. This peak is however likely poorly constrained due to absence of observations at higher temperatures. The peak subsequently decreased to \logt\,$\approx$\,7.1 and 7.4 during the gradual phase.
\end{enumerate}

Our results show that the departures from the Maxwellian distribution can be determined using flare lines observed by SDO/EVE. Furthermore, these departures from Maxwellian can be extreme during the early and impulsive phases of the flare. As we have shown, the non-Maxwellian distributions influence both the temperature and DEM diagnostics of the flare plasma. We suggest that these effects ought to be taken into account during such analyses of flare observations, performed by a number of authors in the past \citep[e.g.,][]{Hannah13,Kennedy13,Cheng14,Sun14,Song15,Gou15,Scullion16,Lee17b}.

\acknowledgments
We thank the anonymous referee for comments that helped to improve the manuscript. We also thank Dr Iain Hannah for making his DEM inversion code public. We acknowledge support from Grants No. 17-16447S and 18-09072S of the Grant Agency of the Czech Republic, as well as institutional support RVO:67985815 from the Czech Academy of Sciences. AIA and EVE data are courtesy of NASA/SDO and the AIA and EVE science teams. 
CHIANTI is a collaborative project involving the NRL (USA), the University of Cambridge (UK), and George Mason University (USA). The ``X-ray Flare'' dataset was prepared by and made available through the NOAA National Geophysical Data Center (NGDC).
\facilities{SDO.}

%
\appendix
\section{EVE lines and their blends}
\label{Appendix:A}

Here, we discuss the details involving individual EVE flare lines used for diagnostics of plasma parameters during the flare, especially $n_\mathrm{e}$, $T$, and $\kappa$.

%
\subsection{\ion{Fe}{18}}
\label{Appendix:A_18}

The EVE line at 93.9\,\AA~is a well-known blend of \ion{Fe}{18} 93.93\,\AA~with \ion{Fe}{20} 93.78\,\AA~\citep[e.g.,][]{ODwyer10,Lemen12,Testa12,Warren12,DelZanna13c}. Unresolvable blends include \ion{Fe}{8}, \ion{Fe}{10}, \ion{Fe}{12}, \ion{Fe}{14}, and other QS and AR lines \citep[e.g.,][]{ODwyer10,DelZanna13b,DelZanna13c} The contribution of these blends is in our case (i.e., for the DEM$_\kappa(T)$ obtained in Sect. \ref{Sect:4.4}) not significant. Resolvable blends in the wings of the \ion{Fe}{18} line include \ion{Ni}{20} 94.50\,\AA~and \ion{Fe}{20} 94.64\,\AA~that were fitted using XCFIT. These blends are typically $<$5\% of 93.9\,\AA~line intensity.

%
\subsection{\ion{Fe}{19}}
\label{Appendix:A_19}

The EVE line at 91\,\AA~is a blend of \ion{Fe}{19} 91.01\,\AA~with \ion{Fe}{21} 91.27\,\AA. \citet{DelZanna13c} lists multiple other unresolvable blends: in quiet Sun (hereafter, QS), \ion{Fe}{10}, \ion{Fe}{11}, and \ion{Fe}{12}, while in active region (AR) conditions additional blends occur from \ion{Fe}{13}, \ion{Fe}{16}, and \ion{O}{8}. We re-calculated the total contribution function including all blends, and verified that for the DEM$_\kappa(T)$ derived for the flare (Sect. \ref{Sect:4.4}) the QS and AR blends at temperatures below \logt\,=\,6.6 are effectively removed by the subtraction of the pre-flare spectrum. The resolvable blends in wings of the 91.0\,\AA~line include \ion{Fe}{20} 90.59\,\AA~and \ion{Ni}{23} 91.87\,\AA, which were approximated using XCFIT. Their intensities are typically $\lesssim$10\% of the 91.0\,\AA~line.

The 424.3\,\AA~EVE line is a blend of \ion{Fe}{19} 424.27\,\AA~with \ion{Fe}{20} 423.93\,\AA. The unresolvable AR blend of \ion{Ar}{15} 423.98\,\AA~(formed at \logt\,$\approx$\,6.7 at Maxwellian conditions) was not included in the theoretical intensity calculations. It contributes less than 10\%. 
The resolved blends include \ion{Fe}{20} 423.11\,\AA, which was included in XCFIT, and \ion{Fe}{19} 425.21\,\AA~line, which was unobserved and thus not fitted.

Finally, the EVE line of \ion{Fe}{19} 108.35\,\AA~is the strongest \ion{Fe}{19} line and thus best for EM analyses (Sect. \ref{Sect:4.4}), as noted by \citet{DelZanna13c}. These authors state that this line is blended with \ion{Fe}{21}, which however contributes only about 10\%. This blend has been neglected, since its contribution is smaller than the overall uncertainty of the line, especially considering the 20\% EVE calibration uncertainty \citep{Woods12}.

%
\subsection{\ion{Fe}{20}}
\label{Appendix:A_20}

The \ion{Fe}{20} 121.85\,\AA~line is a strong line suitable for EM analyses \citep{DelZanna13c}. It has a selfblend at 121.99\,\AA, whose contribution is $\leq$1\% of the main line. The \ion{Fe}{21} 121.21\,\AA~blend, which broadens the line, was included as an additional Gaussian in XCFIT. Its typical contribution to the total intensity is about 8\% of the main line. Several resolved lines nearby include \ion{Fe}{20} 119.98\,\AA~and \ion{Fe}{21} 123.83\,\AA, which were fitted by XCFIT. An unknown QS blend at 122.5\,\AA, possibly \ion{Ne}{6}, was mentioned by \citet{DelZanna13c}. 

%
\subsection{\ion{Fe}{21}}
\label{Appendix:A_21}

Both \ion{Fe}{21} 128.75\,\AA~and 145.73\,\AA~lines used for diagnostics of $n_\mathrm{e}$ are well observed by EVE, as already mentioned in Sect. \ref{Sect:4.1}. The \ion{Fe}{21} 128.75\,\AA~is relatively free of blends, with only a few QS blends \citep{DelZanna13c}. These QS blends are expected to be negligible in the pre-flare subtracted spectra. We have verified this by including their theoretical contribution as a function of $T$ into the synthetic $G_{X,ji}(T,n_\mathrm{e},\kappa)$ and folding over the DEM$_\kappa(T)$ calculated from the observations (see Sect. \ref{Sect:4.4}). The nearby lines, \ion{Fe}{20} 127.84\,\AA~and \ion{Mn}{19} 130.58\,\AA~are resolved in EVE spectra and were included in the fitting with XCFIT.

The \ion{Fe}{21} 145.73\,\AA~is blended by \ion{Mn}{21} 145.4590\,\AA. This blend cannot be subtracted during line profile fitting. However, its contribution to the \ion{Fe}{21} 145.73\,\AA~ intensity is $\leq$10\%. The contribution of this blend is not included in our theoretical calculations of the \ion{Fe}{21} 145.73\,\AA~line. An additional blend could be a \ion{Ni}{10} QS line \citep{DelZanna13c}, not included in CHIANTI v7.1 or v8. Its contribution should however be negligible, since it is a QS line. The nearby lines of \ion{Ni}{22} 144.81\,\AA~and \ion{Fe}{23} 147.25\,\AA~are resolved and were included in the fitting with XCFIT together with their blends.

%
\subsection{\ion{Fe}{22}}
\label{Appendix:A_22}

The \ion{Fe}{22} 135.79\,\AA~line (used for diagnostics of $T$ in Sect. \ref{Sect:4.2}) is a self-blend with the 136.0\,\AA~transition. This line has no other significant blends. Resolved lines nearby include \ion{Fe}{22} 134.69\,\AA~and \ion{Fe}{23} 136.53\,\AA, which were included in the XCFIT together with their blends.

The \ion{Fe}{22} line at 114.4\,\AA~used for diagnostics of $\kappa$ (Sect. \ref{Sect:4.3}) is visible at densities above \logne\,$\approx$\,11.5 \citep{DelZanna13c}. This line does not have any significant unresolved blends. Resolved lines in its vicinity arise from \ion{Fe}{20} at 113.35\,\AA, \ion{Fe}{19} at 115.40\,\AA, and \ion{Mn}{18} at 115.37\,\AA. These lines were included in the Gaussian fitting of the spectra by XCFIT. The QS blends of \ion{Fe}{9}, \ion{Fe}{11}, \ion{Ne}{6} are not significant, and we verified that these are effectively suppressed by the DEM$_\kappa(T)$ obtained in Sect. \ref{Sect:4.4}.

The \ion{Fe}{22} line at 247.2\,\AA~used for diagnostics of $\kappa$ is even weaker than the previous one. It is a sum of \ion{Fe}{22} 247.19\,\AA~with \ion{Fe}{21} 246.95\,\AA. These contributions were summed together in theoretical calculations. The unresolved blend from \ion{S}{11} 246.90\,\AA~(formed at about \logt\,=\,6.3 for Maxwellian distribution) was not included. For the DEM$_\kappa(T)$ obtained here, it contributes about 10\% to the total observed 247\,\AA~line intensity for coronal abundances of \citet{Feldman92}. Strongest resolved lines in the vicinity include \ion{Fe}{13} 246.21\,\AA, \ion{Si}{6} 246.00\,\AA, and \ion{O}{5} 248.46\,\AA, which can be visible in the subtracted spectrum, and were subsequently fitted by XCFIT.

%
\subsection{\ion{Fe}{23}}
\label{Appendix:A_23}

The \ion{Fe}{23} 132.91\,\AA~used for diagnostics of $T$ (Sect. \ref{Sect:4.2}) is blended with \ion{Fe}{20} 132.84\,\AA. This blend has been removed using the procedure outlined by \citet{DelZanna13c}, i.e., from the \ion{Fe}{20} 121.84\,\AA, since both these \ion{Fe}{21} lines are decays to the ground state. The additional blend of \ion{Fe}{19} 132.62\,\AA~is very weak, below 1\%.

The \ion{Fe}{23} 263.77\,\AA~line used for DEM analysis (Sect. \ref{Sect:4.4}) has no significant blends.

%
\subsection{\ion{Fe}{24}}
\label{Appendix:A_24}

The \ion{Fe}{24} 192.03\,\AA~line used for temperature diagnostics (Sect. \ref{Sect:4.2}) is a well-known flare line also observed by Hinode/EIS and other instruments \citep[see, e.g.,][]{Warren01,ODwyer10,ODwyer11,Hara11,Doschek13,Young13,Graham13,Lee17a}. This line has no significant blends, except the QS \ion{O}{5}, \ion{Fe}{11}, and \ion{Fe}{12} \citep{Ko09}, that are negligible in large flares \citep{DelZanna13c}. Nearby resolved lines of \ion{Fe}{12} 191.05\,\AA~and \ion{Ca}{17} 192.85\,\AA~were included in the approximation by XCFIT together with their respective blends.

The \ion{Fe}{24} 255.11\,\AA~line is blended with \ion{Fe}{17} 254.89\,\AA, which contributes about 5\% for the DEMs obtained in Sect. \ref{Sect:4.4}.

%
\subsection{Additional lines used for DEM analyses}
\label{Appendix:A_DEM}

Several lines formed at QS or AR temperatures are used for DEM analyses in Sect. \ref{Sect:4.4}. The \ion{Fe}{10} 177.2\,\AA~is blended with \ion{Fe}{9} 176.96\,\AA, which contributes about 20\%. An additional blend of \ion{Fe}{9} 177.6\,\AA~is weaker by a factor of 2--4, depending on $\kappa$ \citep{Dudik14b}.

The \ion{Fe}{13} 202.04\,\AA~EVE line is blended with \ion{Fe}{12} 201.74\,\AA, which contributes about 15 \%. The \ion{Fe}{14} 211.32\,\AA~has no significant blends. The \ion{Fe}{15} 284.16\,\AA~line is blended in EVE spectra with \ion{Fe}{17} 283.95\,\AA, which contributes about 5\% to the total intensity. Finally, the \ion{Fe}{16} 335.41\,\AA~is blended with \ion{Fe}{21} 335.62\,\AA, which contributes about 5\% of the total intensity.

\bibliographystyle{aasjournal}
\bibliography{2017_EVE_Diag}

\begin{thebibliography}{}
\expandafter\ifx\csname natexlab\endcsname\relax\def\natexlab#1{#1}\fi
\providecommand{\url}[1]{\href{#1}{#1}}

\bibitem[{{Ajello} {et~al.}(2014){Ajello}, {Albert}, {Allafort}, {Baldini},
  {Barbiellini}, {Bastieri}, {Bellazzini}, {Bissaldi}, {Bonamente}, {Brandt},
  {Bregeon}, {Brigida}, {Bruel}, {Buehler}, {Buson}, {Caliandro}, {Cameron},
  {Caraveo}, {Cecchi}, {Charles}, {Chekhtman}, {Chiang}, {Chiaro}, {Ciprini},
  {Claus}, {Cohen-Tanugi}, {Cominsky}, {Conrad}, {Cutini}, {D'Ammando}, {de
  Palma}, {Dermer}, {Desiante}, {Digel}, {Silva}, {Drell}, {Drlica-Wagner},
  {Favuzzi}, {Focke}, {Franckowiak}, {Fukazawa}, {Fusco}, {Gargano},
  {Gasparrini}, {Germani}, {Giglietto}, {Giommi}, {Giordano}, {Giroletti},
  {Glanzman}, {Godfrey}, {Grenier}, {Grove}, {Guiriec}, {Hadasch}, {Hayashida},
  {Hays}, {Horan}, {Hou}, {Hughes}, {Inoue}, {Jackson}, {Jogler},
  {J{\'o}hannesson}, {Johnson}, {Johnson}, {Kamae}, {Kn{\"o}dlseder},
  {Kocevski}, {Kuss}, {Lande}, {Larsson}, {Latronico}, {Longo}, {Loparco},
  {Lott}, {Lovellette}, {Lubrano}, {Mayer}, {Mazziotta}, {McEnery},
  {Michelson}, {Mizuno}, {Moiseev}, {Monte}, {Monzani}, {Morselli},
  {Moskalenko}, {Murgia}, {Murphy}, {Nakamori}, {Nemmen}, {Nuss}, {Ohno},
  {Ohsugi}, {Omodei}, {Orienti}, {Orlando}, {Ormes}, {Paneque}, {Panetta},
  {Perkins}, {Pesce-Rollins}, {Petrosian}, {Piron}, {Pivato}, {Porter},
  {Rain{\`o}}, {Rando}, {Razzano}, {Reimer}, {Reimer}, {Roth}, {Schulz},
  {Sgr{\`o}}, {Siskind}, {Spandre}, {Spinelli}, {Takahashi}, {Thayer},
  {Thayer}, {Thompson}, {Tibaldo}, {Tinivella}, {Tosti}, {Troja}, {Usher},
  {Vandenbroucke}, {Vasileiou}, {Vianello}, {Vitale}, {Werner}, {Winer},
  {Wood}, {Wood}, \& {Yang}}]{Ajello14}
{Ajello}, M., {Albert}, A., {Allafort}, A., {et~al.} 2014, \apj, 789, 20

\bibitem[{{Altun} {et~al.}(2005){Altun}, {Yumak}, {Badnell}, {Colgan}, \&
  {Pindzola}}]{Altun05}
{Altun}, Z., {Yumak}, A., {Badnell}, N.~R., {Colgan}, J., \& {Pindzola}, M.~S.
  2005, \aap, 433, 395

\bibitem[{{Altun} {et~al.}(2006){Altun}, {Yumak}, {Badnell}, {Loch}, \&
  {Pindzola}}]{Altun06}
{Altun}, Z., {Yumak}, A., {Badnell}, N.~R., {Loch}, S.~D., \& {Pindzola}, M.~S.
  2006, \aap, 447, 1165

\bibitem[{{Altun} {et~al.}(2007){Altun}, {Yumak}, {Yavuz}, {Badnell}, {Loch},
  \& {Pindzola}}]{Altun07}
{Altun}, Z., {Yumak}, A., {Yavuz}, I., {et~al.} 2007, \aap, 474, 1051

\bibitem[{{Aulanier} {et~al.}(2012){Aulanier}, {Janvier}, \&
  {Schmieder}}]{Aulanier12}
{Aulanier}, G., {Janvier}, M., \& {Schmieder}, B. 2012, \aap, 543, A110

\bibitem[{{Badnell}(2006)}]{Badnell06a}
{Badnell}, N.~R. 2006, \aap, 447, 389

\bibitem[{{Badnell}(2011)}]{Badnell11}
---. 2011, Computer Physics Communications, 182, 1528

\bibitem[{{Badnell} \& {Griffin}(2001)}]{Badnell01a}
{Badnell}, N.~R., \& {Griffin}, D.~C. 2001, Journal of Physics B Atomic
  Molecular Physics, 34, 681

\bibitem[{{Badnell} {et~al.}(2001){Badnell}, {Griffin}, \&
  {Mitnik}}]{Badnell01b}
{Badnell}, N.~R., {Griffin}, D.~C., \& {Mitnik}, D.~M. 2001, Journal of Physics
  B Atomic Molecular Physics, 34, 5071

\bibitem[{{Badnell} {et~al.}(2003){Badnell}, {O'Mullane}, {Summers}, {Altun},
  {Bautista}, {Colgan}, {Gorczyca}, {Mitnik}, {Pindzola}, \&
  {Zatsarinny}}]{Badnell03}
{Badnell}, N.~R., {O'Mullane}, M.~G., {Summers}, H.~P., {et~al.} 2003, \aap,
  406, 1151

\bibitem[{{Battaglia} \& {Kontar}(2013)}]{Battaglia13}
{Battaglia}, M., \& {Kontar}, E.~P. 2013, \apj, 779, 107

\bibitem[{{Battaglia} {et~al.}(2015){Battaglia}, {Motorina}, \&
  {Kontar}}]{Battaglia15}
{Battaglia}, M., {Motorina}, G., \& {Kontar}, E.~P. 2015, \apj, 815, 73

\bibitem[{{Bautista} \& {Badnell}(2007)}]{Bautista07}
{Bautista}, M.~A., \& {Badnell}, N.~R. 2007, \aap, 466, 755

\bibitem[{{Berrington} \& {Tully}(1997)}]{Berrington97}
{Berrington}, K.~A., \& {Tully}, J.~A. 1997, \aaps, 126,
  doi:10.1051/aas:1997384

\bibitem[{{Bian} {et~al.}(2014){Bian}, {Emslie}, {Stackhouse}, \&
  {Kontar}}]{Bian14}
{Bian}, N.~H., {Emslie}, A.~G., {Stackhouse}, D.~J., \& {Kontar}, E.~P. 2014,
  \apj, 796, 142

\bibitem[{{Boerner} {et~al.}(2012){Boerner}, {Edwards}, {Lemen}, {Rausch},
  {Schrijver}, {Shine}, {Shing}, {Stern}, {Tarbell}, {Title}, {Wolfson},
  {Soufli}, {Spiller}, {Gullikson}, {McKenzie}, {Windt}, {Golub}, {Podgorski},
  {Testa}, \& {Weber}}]{Boerner12}
{Boerner}, P., {Edwards}, C., {Lemen}, J., {et~al.} 2012, \solphys, 275, 41

\bibitem[{{Bradshaw} {et~al.}(2004){Bradshaw}, {Del Zanna}, \&
  {Mason}}]{Bradshaw04}
{Bradshaw}, S.~J., {Del Zanna}, G., \& {Mason}, H.~E. 2004, \aap, 425, 287

\bibitem[{{Brown}(1971)}]{Brown71}
{Brown}, J.~C. 1971, \solphys, 18, 489

\bibitem[{{Brown} {et~al.}(2016){Brown}, {Fletcher}, \& {Labrosse}}]{Brown16}
{Brown}, S.~A., {Fletcher}, L., \& {Labrosse}, N. 2016, \aap, 596, A51

\bibitem[{{Bykov} {et~al.}(2013){Bykov}, {Malkov}, {Raymond},
  {Krassilchtchikov}, \& {Vladimirov}}]{Bykov13}
{Bykov}, A.~M., {Malkov}, M.~A., {Raymond}, J.~C., {Krassilchtchikov}, A.~M.,
  \& {Vladimirov}, A.~E. 2013, \ssr, 178, 599

\bibitem[{{Cheng} {et~al.}(2014){Cheng}, {Ding}, {Zhang}, {Srivastava}, {Guo},
  {Chen}, \& {Sun}}]{Cheng14}
{Cheng}, X., {Ding}, M.~D., {Zhang}, J., {et~al.} 2014, \apjl, 789, L35

\bibitem[{{Chidichimo} {et~al.}(2005){Chidichimo}, {Del Zanna}, {Mason},
  {Badnell}, {Tully}, \& {Berrington}}]{Chidichimo05}
{Chidichimo}, M.~C., {Del Zanna}, G., {Mason}, H.~E., {et~al.} 2005, \aap, 430,
  331

\bibitem[{{Chintzoglou} {et~al.}(2015){Chintzoglou}, {Patsourakos}, \&
  {Vourlidas}}]{Chintzoglou15}
{Chintzoglou}, G., {Patsourakos}, S., \& {Vourlidas}, A. 2015, \apj, 809, 34

\bibitem[{{Colgan} {et~al.}(2004){Colgan}, {Pindzola}, \& {Badnell}}]{Colgan04}
{Colgan}, J., {Pindzola}, M.~S., \& {Badnell}, N.~R. 2004, \aap, 417, 1183

\bibitem[{{Colgan} {et~al.}(2003){Colgan}, {Pindzola}, {Whiteford}, \&
  {Badnell}}]{Colgan03}
{Colgan}, J., {Pindzola}, M.~S., {Whiteford}, A.~D., \& {Badnell}, N.~R. 2003,
  \aap, 412, 597

\bibitem[{{Culhane} {et~al.}(2007){Culhane}, {Harra}, {James}, {Al-Janabi},
  {Bradley}, {Chaudry}, {Rees}, {Tandy}, {Thomas}, {Whillock}, {Winter},
  {Doschek}, {Korendyke}, {Brown}, {Myers}, {Mariska}, {Seely}, {Lang}, {Kent},
  {Shaughnessy}, {Young}, {Simnett}, {Castelli}, {Mahmoud}, {Mapson-Menard},
  {Probyn}, {Thomas}, {Davila}, {Dere}, {Windt}, {Shea}, {Hagood}, {Moye},
  {Hara}, {Watanabe}, {Matsuzaki}, {Kosugi}, {Hansteen}, \&
  {Wikstol}}]{Culhane07}
{Culhane}, J.~L., {Harra}, L.~K., {James}, A.~M., {et~al.} 2007, \solphys, 243,
  19

\bibitem[{{De Pontieu} {et~al.}(2014){De Pontieu}, {Title}, {Lemen}, {Kushner},
  {Akin}, {Allard}, {Berger}, {Boerner}, {Cheung}, {Chou}, {Drake}, {Duncan},
  {Freeland}, {Heyman}, {Hoffman}, {Hurlburt}, {Lindgren}, {Mathur}, {Rehse},
  {Sabolish}, {Seguin}, {Schrijver}, {Tarbell}, {W{\"u}lser}, {Wolfson},
  {Yanari}, {Mudge}, {Nguyen-Phuc}, {Timmons}, {van Bezooijen}, {Weingrod},
  {Brookner}, {Butcher}, {Dougherty}, {Eder}, {Knagenhjelm}, {Larsen},
  {Mansir}, {Phan}, {Boyle}, {Cheimets}, {DeLuca}, {Golub}, {Gates}, {Hertz},
  {McKillop}, {Park}, {Perry}, {Podgorski}, {Reeves}, {Saar}, {Testa}, {Tian},
  {Weber}, {Dunn}, {Eccles}, {Jaeggli}, {Kankelborg}, {Mashburn}, {Pust},
  {Springer}, {Carvalho}, {Kleint}, {Marmie}, {Mazmanian}, {Pereira}, {Sawyer},
  {Strong}, {Worden}, {Carlsson}, {Hansteen}, {Leenaarts}, {Wiesmann},
  {Aloise}, {Chu}, {Bush}, {Scherrer}, {Brekke}, {Martinez-Sykora}, {Lites},
  {McIntosh}, {Uitenbroek}, {Okamoto}, {Gummin}, {Auker}, {Jerram}, {Pool}, \&
  {Waltham}}]{DePontieu14}
{De Pontieu}, B., {Title}, A.~M., {Lemen}, J.~R., {et~al.} 2014, \solphys, 289,
  2733

\bibitem[{{Del Zanna}(2006)}]{DelZanna06}
{Del Zanna}, G. 2006, \aap, 459, 307

\bibitem[{{Del Zanna}(2013)}]{DelZanna13b}
---. 2013, \aap, 558, A73

\bibitem[{{Del Zanna} {et~al.}(2005){Del Zanna}, {Chidichimo}, \&
  {Mason}}]{DelZanna05}
{Del Zanna}, G., {Chidichimo}, M.~C., \& {Mason}, H.~E. 2005, \aap, 432, 1137

\bibitem[{{Del Zanna} {et~al.}(2015){Del Zanna}, {Dere}, {Young}, {Landi}, \&
  {Mason}}]{DelZanna15b}
{Del Zanna}, G., {Dere}, K.~P., {Young}, P.~R., {Landi}, E., \& {Mason}, H.~E.
  2015, \aap, 582, A56

\bibitem[{{Del Zanna} \& {Mason}(2003)}]{DelZanna03}
{Del Zanna}, G., \& {Mason}, H.~E. 2003, \aap, 406, 1089

\bibitem[{{Del Zanna} \& {Woods}(2013)}]{DelZanna13c}
{Del Zanna}, G., \& {Woods}, T.~N. 2013, \aap, 555, A59

\bibitem[{{Dere}(2007)}]{Dere07}
{Dere}, K.~P. 2007, \aap, 466, 771

\bibitem[{{Dere} {et~al.}(1997){Dere}, {Landi}, {Mason}, {Monsignori Fossi}, \&
  {Young}}]{Dere97}
{Dere}, K.~P., {Landi}, E., {Mason}, H.~E., {Monsignori Fossi}, B.~C., \&
  {Young}, P.~R. 1997, \aaps, 125, 149

\bibitem[{{Dere} {et~al.}(2009){Dere}, {Landi}, {Young}, {Del Zanna},
  {Landini}, \& {Mason}}]{Dere09}
{Dere}, K.~P., {Landi}, E., {Young}, P.~R., {et~al.} 2009, \aap, 498, 915

\bibitem[{{Doschek} {et~al.}(1979){Doschek}, {Kreplin}, \&
  {Feldman}}]{Doschek79}
{Doschek}, G.~A., {Kreplin}, R.~W., \& {Feldman}, U. 1979, \apjl, 233, L157

\bibitem[{{Doschek} \& {Tanaka}(1987)}]{Doschek87}
{Doschek}, G.~A., \& {Tanaka}, K. 1987, \apj, 323, 799

\bibitem[{{Doschek} \& {Warren}(2016)}]{Doschek16}
{Doschek}, G.~A., \& {Warren}, H.~P. 2016, \apj, 825, 36

\bibitem[{{Doschek} {et~al.}(2015){Doschek}, {Warren}, \&
  {Feldman}}]{Doschek15}
{Doschek}, G.~A., {Warren}, H.~P., \& {Feldman}, U. 2015, \apjl, 808, L7

\bibitem[{{Doschek} {et~al.}(2013){Doschek}, {Warren}, \& {Young}}]{Doschek13}
{Doschek}, G.~A., {Warren}, H.~P., \& {Young}, P.~R. 2013, \apj, 767, 55

\bibitem[{{Dud{\'{\i}}k} {et~al.}(2014{\natexlab{a}}){Dud{\'{\i}}k}, {Del
  Zanna}, {Dzif{\v c}{\'a}kov{\'a}}, {Mason}, \& {Golub}}]{Dudik14a}
{Dud{\'{\i}}k}, J., {Del Zanna}, G., {Dzif{\v c}{\'a}kov{\'a}}, E., {Mason},
  H.~E., \& {Golub}, L. 2014{\natexlab{a}}, \apjl, 780, L12

\bibitem[{{Dud{\'{\i}}k} {et~al.}(2014{\natexlab{b}}){Dud{\'{\i}}k}, {Del
  Zanna}, {Mason}, \& {Dzif{\v c}{\'a}kov{\'a}}}]{Dudik14b}
{Dud{\'{\i}}k}, J., {Del Zanna}, G., {Mason}, H.~E., \& {Dzif{\v
  c}{\'a}kov{\'a}}, E. 2014{\natexlab{b}}, \aap, 570, A124

\bibitem[{{Dud{\'{\i}}k} {et~al.}(2012){Dud{\'{\i}}k}, {Ka{\v s}parov{\'a}},
  {Dzif{\v c}{\'a}kov{\'a}}, {Karlick{\'y}}, \& {Mackovjak}}]{Dudik12}
{Dud{\'{\i}}k}, J., {Ka{\v s}parov{\'a}}, J., {Dzif{\v c}{\'a}kov{\'a}}, E.,
  {Karlick{\'y}}, M., \& {Mackovjak}, {\v S}. 2012, \aap, 539, A107

\bibitem[{{Dud{\'{\i}}k} {et~al.}(2017{\natexlab{a}}){Dud{\'{\i}}k}, {Polito},
  {Dzif{\v c}{\'a}kov{\'a}}, {Del Zanna}, \& {Testa}}]{Dudik17b}
{Dud{\'{\i}}k}, J., {Polito}, V., {Dzif{\v c}{\'a}kov{\'a}}, E., {Del Zanna},
  G., \& {Testa}, P. 2017{\natexlab{a}}, \apj, 842, 19

\bibitem[{{Dud{\'{\i}}k} {et~al.}(2017{\natexlab{b}}){Dud{\'{\i}}k},
  {Zuccarello}, {Aulanier}, {Schmieder}, \& {D{\'e}moulin}}]{Dudik17c}
{Dud{\'{\i}}k}, J., {Zuccarello}, F.~P., {Aulanier}, G., {Schmieder}, B., \&
  {D{\'e}moulin}, P. 2017{\natexlab{b}}, \apj, 844, 54

\bibitem[{{Dud{\'{\i}}k} {et~al.}(2015){Dud{\'{\i}}k}, {Mackovjak}, {Dzif{\v
  c}{\'a}kov{\'a}}, {Del Zanna}, {Williams}, {Karlick{\'y}}, {Mason},
  {L{\"o}rin{\v c}{\'{\i}}k}, {Kotr{\v c}}, {F{\'a}rn{\'{\i}}k}, \&
  {Zemanov{\'a}}}]{Dudik15}
{Dud{\'{\i}}k}, J., {Mackovjak}, {\v S}., {Dzif{\v c}{\'a}kov{\'a}}, E.,
  {et~al.} 2015, \apj, 807, 123

\bibitem[{{Dud{\'{\i}}k} {et~al.}(2017{\natexlab{c}}){Dud{\'{\i}}k}, {Dzif{\v
  c}{\'a}kov{\'a}}, {Meyer-Vernet}, {Del Zanna}, {Young}, {Giunta},
  {Sylwester}, {Sylwester}, {Oka}, {Mason}, {Vocks}, {Matteini}, {Krucker},
  {Williams}, \& {Mackovjak}}]{Dudik17a}
{Dud{\'{\i}}k}, J., {Dzif{\v c}{\'a}kov{\'a}}, E., {Meyer-Vernet}, N., {et~al.}
  2017{\natexlab{c}}, \solphys, 292, 100

\bibitem[{{Dungey}(153)}]{Dungey53}
{Dungey}, J.~W. 153, Phil. Mag., 44

\bibitem[{{Dzif{\v c}{\'a}kov{\'a}}(2002)}]{Dzifcakova02}
{Dzif{\v c}{\'a}kov{\'a}}, E. 2002, \solphys, 208, 91

\bibitem[{{Dzif{\v c}{\'a}kov{\'a}}(2006)}]{Dzifcakova06}
---. 2006, \solphys, 234, 243

\bibitem[{{Dzif{\v c}{\'a}kov{\'a}} \& {Dud{\'{\i}}k}(2013)}]{Dzifcakova13a}
{Dzif{\v c}{\'a}kov{\'a}}, E., \& {Dud{\'{\i}}k}, J. 2013, \apjs, 206, 6

\bibitem[{{Dzif{\v c}{\'a}kov{\'a}} {et~al.}(2015){Dzif{\v c}{\'a}kov{\'a}},
  {Dud{\'{\i}}k}, {Kotr{\v c}}, {F{\'a}rn{\'{\i}}k}, \&
  {Zemanov{\'a}}}]{Dzifcakova15}
{Dzif{\v c}{\'a}kov{\'a}}, E., {Dud{\'{\i}}k}, J., {Kotr{\v c}}, P.,
  {F{\'a}rn{\'{\i}}k}, F., \& {Zemanov{\'a}}, A. 2015, \apjs, 217, 14

\bibitem[{{Dzif{\v c}{\'a}kov{\'a}} \& {Kulinov{\'a}}(2010)}]{Dzifcakova10}
{Dzif{\v c}{\'a}kov{\'a}}, E., \& {Kulinov{\'a}}, A. 2010, \solphys, 263, 25

\bibitem[{{Dzif\v{c}{\'a}kov{\'a}}(1992)}]{Dzifcakova92}
{Dzif\v{c}{\'a}kov{\'a}}, E. 1992, \solphys, 140, 247

\bibitem[{{Emslie} {et~al.}(2012){Emslie}, {Dennis}, {Shih}, {Chamberlin},
  {Mewaldt}, {Moore}, {Share}, {Vourlidas}, \& {Welsch}}]{Emslie12}
{Emslie}, A.~G., {Dennis}, B.~R., {Shih}, A.~Y., {et~al.} 2012, \apj, 759, 71

\bibitem[{{Feldman} {et~al.}(1992){Feldman}, {Mandelbaum}, {Seely}, {Doschek},
  \& {Gursky}}]{Feldman92}
{Feldman}, U., {Mandelbaum}, P., {Seely}, J.~F., {Doschek}, G.~A., \& {Gursky},
  H. 1992, \apjs, 81, 387

\bibitem[{{Fletcher} {et~al.}(2011){Fletcher}, {Dennis}, {Hudson}, {Krucker},
  {Phillips}, {Veronig}, {Battaglia}, {Bone}, {Caspi}, {Chen}, {Gallagher},
  {Grigis}, {Ji}, {Liu}, {Milligan}, \& {Temmer}}]{Fletcher11}
{Fletcher}, L., {Dennis}, B.~R., {Hudson}, H.~S., {et~al.} 2011, \ssr, 159, 19

\bibitem[{{Glesener} {et~al.}(2017){Glesener}, {Krucker}, {Hannah}, {Hudson},
  {Grefenstette}, {White}, {Smith}, \& {Marsh}}]{Glesener17}
{Glesener}, L., {Krucker}, S., {Hannah}, I.~G., {et~al.} 2017, \apj, 845, 122

\bibitem[{{Golub} {et~al.}(1989){Golub}, {Hartquist}, \& {Quillen}}]{Golub89}
{Golub}, L., {Hartquist}, T.~W., \& {Quillen}, A.~C. 1989, \solphys, 122, 245

\bibitem[{{Golub} {et~al.}(2007){Golub}, {Deluca}, {Austin}, {Bookbinder},
  {Caldwell}, {Cheimets}, {Cirtain}, {Cosmo}, {Reid}, {Sette}, {Weber},
  {Sakao}, {Kano}, {Shibasaki}, {Hara}, {Tsuneta}, {Kumagai}, {Tamura},
  {Shimojo}, {McCracken}, {Carpenter}, {Haight}, {Siler}, {Wright}, {Tucker},
  {Rutledge}, {Barbera}, {Peres}, \& {Varisco}}]{Golub07}
{Golub}, L., {Deluca}, E., {Austin}, G., {et~al.} 2007, \solphys, 243, 63

\bibitem[{{Gou} {et~al.}(2015){Gou}, {Liu}, \& {Wang}}]{Gou15}
{Gou}, T., {Liu}, R., \& {Wang}, Y. 2015, \solphys, 290, 2211

\bibitem[{{Graham} {et~al.}(2013){Graham}, {Hannah}, {Fletcher}, \&
  {Milligan}}]{Graham13}
{Graham}, D.~R., {Hannah}, I.~G., {Fletcher}, L., \& {Milligan}, R.~O. 2013,
  \apj, 767, 83

\bibitem[{{Gu}(2003)}]{Gu03}
{Gu}, M.~F. 2003, \apj, 582, 1241

\bibitem[{{Hahn} \& {Savin}(2015)}]{Hahn15b}
{Hahn}, M., \& {Savin}, D.~W. 2015, \apj, 809, 178

\bibitem[{{Hannah} {et~al.}(2008){Hannah}, {Christe}, {Krucker}, {Hurford},
  {Hudson}, \& {Lin}}]{Hannah08}
{Hannah}, I.~G., {Christe}, S., {Krucker}, S., {et~al.} 2008, \apj, 677, 704

\bibitem[{{Hannah} \& {Kontar}(2012)}]{Hannah12}
{Hannah}, I.~G., \& {Kontar}, E.~P. 2012, \aap, 539, A146

\bibitem[{{Hannah} \& {Kontar}(2013)}]{Hannah13}
---. 2013, \aap, 553, A10

\bibitem[{{Hara} {et~al.}(2011){Hara}, {Watanabe}, {Harra}, {Culhane}, \&
  {Young}}]{Hara11}
{Hara}, H., {Watanabe}, T., {Harra}, L.~K., {Culhane}, J.~L., \& {Young}, P.~R.
  2011, \apj, 741, 107

\bibitem[{{Harra} {et~al.}(2016){Harra}, {Schrijver}, {Janvier}, {Toriumi},
  {Hudson}, {Matthews}, {Woods}, {Hara}, {Guedel}, {Kowalski}, {Osten},
  {Kusano}, \& {Lueftinger}}]{Harra16}
{Harra}, L.~K., {Schrijver}, C.~J., {Janvier}, M., {et~al.} 2016, \solphys,
  291, 1761

\bibitem[{{Hasegawa} {et~al.}(1985){Hasegawa}, {Mima}, \&
  {Duong-van}}]{Hasegawa85}
{Hasegawa}, A., {Mima}, K., \& {Duong-van}, M. 1985, Physical Review Letters,
  54, 2608

\bibitem[{{Holman} {et~al.}(2003){Holman}, {Sui}, {Schwartz}, \&
  {Emslie}}]{Holman03}
{Holman}, G.~D., {Sui}, L., {Schwartz}, R.~A., \& {Emslie}, A.~G. 2003, \apjl,
  595, L97

\bibitem[{{Holman} {et~al.}(2011){Holman}, {Aschwanden}, {Aurass}, {Battaglia},
  {Grigis}, {Kontar}, {Liu}, {Saint-Hilaire}, \& {Zharkova}}]{Holman11}
{Holman}, G.~D., {Aschwanden}, M.~J., {Aurass}, H., {et~al.} 2011, \ssr, 260

\bibitem[{{Janvier}(2017)}]{Janvier17}
{Janvier}, M. 2017, Journal of Plasma Physics, 83, 535830101

\bibitem[{{Janvier} {et~al.}(2015){Janvier}, {Aulanier}, \&
  {D{\'e}moulin}}]{Janvier15}
{Janvier}, M., {Aulanier}, G., \& {D{\'e}moulin}, P. 2015, \solphys, 290, 3425

\bibitem[{{Janvier} {et~al.}(2013){Janvier}, {Aulanier}, {Pariat}, \&
  {D{\'e}moulin}}]{Janvier13}
{Janvier}, M., {Aulanier}, G., {Pariat}, E., \& {D{\'e}moulin}, P. 2013, \aap,
  555, A77

\bibitem[{{Jeffrey} {et~al.}(2016){Jeffrey}, {Fletcher}, \&
  {Labrosse}}]{Jeffrey16}
{Jeffrey}, N.~L.~S., {Fletcher}, L., \& {Labrosse}, N. 2016, \aap, 590, A99

\bibitem[{{Jeffrey} {et~al.}(2017){Jeffrey}, {Fletcher}, \&
  {Labrosse}}]{Jeffrey17}
---. 2017, \apj, 836, 35

\bibitem[{{Ka{\v s}parov{\'a}} \& {Karlick{\'y}}(2009)}]{Kasparova09b}
{Ka{\v s}parov{\'a}}, J., \& {Karlick{\'y}}, M. 2009, \aap, 497, L13

\bibitem[{{Kennedy} {et~al.}(2013){Kennedy}, {Milligan}, {Mathioudakis}, \&
  {Keenan}}]{Kennedy13}
{Kennedy}, M.~B., {Milligan}, R.~O., {Mathioudakis}, M., \& {Keenan}, F.~P.
  2013, \apj, 779, 84

\bibitem[{{Ko} {et~al.}(2009){Ko}, {Doschek}, {Warren}, \& {Young}}]{Ko09}
{Ko}, Y.-K., {Doschek}, G.~A., {Warren}, H.~P., \& {Young}, P.~R. 2009, \apj,
  697, 1956

\bibitem[{{Kontar} {et~al.}(2011){Kontar}, {Brown}, {Emslie}, {Hajdas},
  {Holman}, {Hurford}, {Ka{\v s}parov{\'a}}, {Mallik}, {Massone}, {McConnell},
  {Piana}, {Prato}, {Schmahl}, \& {Suarez-Garcia}}]{Kontar11}
{Kontar}, E.~P., {Brown}, J.~C., {Emslie}, A.~G., {et~al.} 2011, \ssr, 279

\bibitem[{{Kouloumvakos} {et~al.}(2016){Kouloumvakos}, {Patsourakos}, {Nindos},
  {Vourlidas}, {Anastasiadis}, {Hillaris}, \& {Sandberg}}]{Kouloumvakos16}
{Kouloumvakos}, A., {Patsourakos}, S., {Nindos}, A., {et~al.} 2016, \apj, 821,
  31

\bibitem[{{Krucker} {et~al.}(2008){Krucker}, {Battaglia}, {Cargill},
  {Fletcher}, {Hudson}, {MacKinnon}, {Masuda}, {Sui}, {Tomczak}, {Veronig},
  {Vlahos}, \& {White}}]{Krucker08b}
{Krucker}, S., {Battaglia}, M., {Cargill}, P.~J., {et~al.} 2008, \aapr, 16, 155

\bibitem[{{Kuhar} {et~al.}(2016){Kuhar}, {Krucker}, {Mart{\'{\i}}nez Oliveros},
  {Battaglia}, {Kleint}, {Casadei}, \& {Hudson}}]{Kuhar16}
{Kuhar}, M., {Krucker}, S., {Mart{\'{\i}}nez Oliveros}, J.~C., {et~al.} 2016,
  \apj, 816, 6

\bibitem[{{Laming} \& {Lepri}(2007)}]{Laming07}
{Laming}, J.~M., \& {Lepri}, S.~T. 2007, \apj, 660, 1642

\bibitem[{{Landi} \& {Gu}(2006)}]{Landi06}
{Landi}, E., \& {Gu}, M.~F. 2006, \apj, 640, 1171

\bibitem[{{Landi} {et~al.}(2013){Landi}, {Young}, {Dere}, {Del Zanna}, \&
  {Mason}}]{Landi13}
{Landi}, E., {Young}, P.~R., {Dere}, K.~P., {Del Zanna}, G., \& {Mason}, H.~E.
  2013, \apj, 763, 86

\bibitem[{{Lee} {et~al.}(2017{\natexlab{a}}){Lee}, {Raymond}, {Reeves}, {Moon},
  \& {Kim}}]{Lee17b}
{Lee}, J.-Y., {Raymond}, J.~C., {Reeves}, K.~K., {Moon}, Y.-J., \& {Kim}, K.-S.
  2017{\natexlab{a}}, \apj, 844, 3

\bibitem[{{Lee} {et~al.}(2017{\natexlab{b}}){Lee}, {Imada}, {Watanabe},
  {Bamba}, \& {Brooks}}]{Lee17a}
{Lee}, K.-S., {Imada}, S., {Watanabe}, K., {Bamba}, Y., \& {Brooks}, D.~H.
  2017{\natexlab{b}}, \apj, 836, 150

\bibitem[{{Lemen} {et~al.}(2012){Lemen}, {Title}, {Akin}, {Boerner}, {Chou},
  {Drake}, {Duncan}, {Edwards}, {Friedlaender}, {Heyman}, {Hurlburt}, {Katz},
  {Kushner}, {Levay}, {Lindgren}, {Mathur}, {McFeaters}, {Mitchell}, {Rehse},
  {Schrijver}, {Springer}, {Stern}, {Tarbell}, {Wuelser}, {Wolfson}, {Yanari},
  {Bookbinder}, {Cheimets}, {Caldwell}, {Deluca}, {Gates}, {Golub}, {Park},
  {Podgorski}, {Bush}, {Scherrer}, {Gummin}, {Smith}, {Auker}, {Jerram},
  {Pool}, {Soufli}, {Windt}, {Beardsley}, {Clapp}, {Lang}, \&
  {Waltham}}]{Lemen12}
{Lemen}, J.~R., {Title}, A.~M., {Akin}, D.~J., {et~al.} 2012, \solphys, 275, 17

\bibitem[{{Lin} \& {Hudson}(1971)}]{Lin71}
{Lin}, R.~P., \& {Hudson}, H.~S. 1971, \solphys, 17, 412

\bibitem[{{Lin} {et~al.}(2002){Lin}, {Dennis}, {Hurford}, {Smith}, {Zehnder},
  {Harvey}, {Curtis}, {Pankow}, {Turin}, {Bester}, {Csillaghy}, {Lewis},
  {Madden}, {van Beek}, {Appleby}, {Raudorf}, {McTiernan}, {Ramaty}, {Schmahl},
  {Schwartz}, {Krucker}, {Abiad}, {Quinn}, {Berg}, {Hashii}, {Sterling},
  {Jackson}, {Pratt}, {Campbell}, {Malone}, {Landis}, {Barrington-Leigh},
  {Slassi-Sennou}, {Cork}, {Clark}, {Amato}, {Orwig}, {Boyle}, {Banks},
  {Shirey}, {Tolbert}, {Zarro}, {Snow}, {Thomsen}, {Henneck}, {McHedlishvili},
  {Ming}, {Fivian}, {Jordan}, {Wanner}, {Crubb}, {Preble}, {Matranga}, {Benz},
  {Hudson}, {Canfield}, {Holman}, {Crannell}, {Kosugi}, {Emslie}, {Vilmer},
  {Brown}, {Johns-Krull}, {Aschwanden}, {Metcalf}, \& {Conway}}]{Lin02}
{Lin}, R.~P., {Dennis}, B.~R., {Hurford}, G.~J., {et~al.} 2002, \solphys, 210,
  3

\bibitem[{{Livadiotis}(2015)}]{Livadiotis15b}
{Livadiotis}, G. 2015, Journal of Geophysical Research (Space Physics), 120,
  1607

\bibitem[{{Mackovjak} {et~al.}(2014){Mackovjak}, {Dzif{\v c}{\'a}kov{\'a}}, \&
  {Dud{\'{\i}}k}}]{Mackovjak14}
{Mackovjak}, {\v S}., {Dzif{\v c}{\'a}kov{\'a}}, E., \& {Dud{\'{\i}}k}, J.
  2014, \aap, 564, A130

\bibitem[{{Mason} {et~al.}(1984){Mason}, {Bhatia}, {Neupert}, {Swartz}, \&
  {Kastner}}]{Mason84}
{Mason}, H.~E., {Bhatia}, A.~K., {Neupert}, W.~M., {Swartz}, M., \& {Kastner},
  S.~O. 1984, \solphys, 92, 199

\bibitem[{{Mason} {et~al.}(1979){Mason}, {Doschek}, {Feldman}, \&
  {Bhatia}}]{Mason79}
{Mason}, H.~E., {Doschek}, G.~A., {Feldman}, U., \& {Bhatia}, A.~K. 1979, \aap,
  73, 74

\bibitem[{{Mason} \& {Monsignori Fossi}(1994)}]{Mason94}
{Mason}, H.~E., \& {Monsignori Fossi}, B.~C. 1994, \aapr, 6, 123

\bibitem[{{Meyer-Vernet}(2007)}]{Meyer-Vernet07}
{Meyer-Vernet}, N. 2007, {Basics of the Solar Wind} (Cambridge University
  Press)

\bibitem[{{Milligan} {et~al.}(2012){Milligan}, {Kennedy}, {Mathioudakis}, \&
  {Keenan}}]{Milligan12}
{Milligan}, R.~O., {Kennedy}, M.~B., {Mathioudakis}, M., \& {Keenan}, F.~P.
  2012, \apjl, 755, L16

\bibitem[{{Mitnik} \& {Badnell}(2004)}]{Mitnik04}
{Mitnik}, D.~M., \& {Badnell}, N.~R. 2004, \aap, 425, 1153

\bibitem[{{Nikoli{\'c}} {et~al.}(2010){Nikoli{\'c}}, {Gorczyca}, {Korista}, \&
  {Badnell}}]{Nikolic10}
{Nikoli{\'c}}, D., {Gorczyca}, T.~W., {Korista}, K.~T., \& {Badnell}, N.~R.
  2010, \aap, 516, A97

\bibitem[{{O'Dwyer} {et~al.}(2011){O'Dwyer}, {Del Zanna}, {Mason}, {Sterling},
  {Tripathi}, \& {Young}}]{ODwyer11}
{O'Dwyer}, B., {Del Zanna}, G., {Mason}, H.~E., {et~al.} 2011, \aap, 525, A137

\bibitem[{{O'Dwyer} {et~al.}(2010){O'Dwyer}, {Del Zanna}, {Mason}, {Weber}, \&
  {Tripathi}}]{ODwyer10}
{O'Dwyer}, B., {Del Zanna}, G., {Mason}, H.~E., {Weber}, M.~A., \& {Tripathi},
  D. 2010, \aap, 521, A21

\bibitem[{{Oka} {et~al.}(2013){Oka}, {Ishikawa}, {Saint-Hilaire}, {Krucker}, \&
  {Lin}}]{Oka13}
{Oka}, M., {Ishikawa}, S., {Saint-Hilaire}, P., {Krucker}, S., \& {Lin}, R.~P.
  2013, \apj, 764, 6

\bibitem[{{Oka} {et~al.}(2015){Oka}, {Krucker}, {Hudson}, \&
  {Saint-Hilaire}}]{Oka15}
{Oka}, M., {Krucker}, S., {Hudson}, H.~S., \& {Saint-Hilaire}, P. 2015, \apj,
  799, 129

\bibitem[{{Olbert}(1968)}]{Olbert68}
{Olbert}, S. 1968, in Astrophysics and Space Science Library, Vol.~10, Physics
  of the Magnetosphere, ed. R.~D.~L. {Carovillano} \& J.~F. {McClay}, 641

\bibitem[{{Owocki} \& {Scudder}(1983)}]{Owocki83}
{Owocki}, S.~P., \& {Scudder}, J.~D. 1983, \apj, 270, 758

\bibitem[{{Parker}(1957)}]{Parker57}
{Parker}, E.~N. 1957, \jgr, 62, 509

\bibitem[{{Patsourakos} {et~al.}(2016){Patsourakos}, {Georgoulis}, {Vourlidas},
  {Nindos}, {Sarris}, {Anagnostopoulos}, {Anastasiadis}, {Chintzoglou},
  {Daglis}, {Gontikakis}, {Hatzigeorgiu}, {Iliopoulos}, {Katsavrias},
  {Kouloumvakos}, {Moraitis}, {Nieves-Chinchilla}, {Pavlos}, {Sarafopoulos},
  {Syntelis}, {Tsironis}, {Tziotziou}, {Vogiatzis}, {Balasis}, {Georgiou},
  {Karakatsanis}, {Malandraki}, {Papadimitriou}, {Odstr{\v c}il}, {Pavlos},
  {Podlachikova}, {Sandberg}, {Turner}, {Xenakis}, {Sarris}, {Tsinganos}, \&
  {Vlahos}}]{Patsourakos16}
{Patsourakos}, S., {Georgoulis}, M.~K., {Vourlidas}, A., {et~al.} 2016, \apj,
  817, 14

\bibitem[{{Pesnell} {et~al.}(2012){Pesnell}, {Thompson}, \&
  {Chamberlin}}]{Pesnell12}
{Pesnell}, W.~D., {Thompson}, B.~J., \& {Chamberlin}, P.~C. 2012, \solphys,
  275, 3

\bibitem[{{Petkaki} {et~al.}(2012){Petkaki}, {Del Zanna}, {Mason}, \&
  {Bradshaw}}]{Petkaki12}
{Petkaki}, P., {Del Zanna}, G., {Mason}, H.~E., \& {Bradshaw}, S.~J. 2012,
  \aap, 547, A25

\bibitem[{{Phillips} {et~al.}(2008){Phillips}, {Feldman}, \&
  {Landi}}]{Phillips08}
{Phillips}, K.~J.~H., {Feldman}, U., \& {Landi}, E. 2008, {Ultraviolet and
  X-ray Spectroscopy of the Solar Atmosphere} (Cambridge University Press)

\bibitem[{{Pierrard} \& {Lazar}(2010)}]{Pierrard10}
{Pierrard}, V., \& {Lazar}, M. 2010, \solphys, 267, 153

\bibitem[{{Polito} {et~al.}(2017){Polito}, {Del Zanna}, {Valori}, {Pariat},
  {Mason}, {Dud{\'{\i}}k}, \& {Janvier}}]{Polito17}
{Polito}, V., {Del Zanna}, G., {Valori}, G., {et~al.} 2017, \aap, 601, A39

\bibitem[{{Polito} {et~al.}(2016){Polito}, {Reep}, {Reeves}, {Sim{\~o}es},
  {Dud{\'{\i}}k}, {Del Zanna}, {Mason}, \& {Golub}}]{Polito16a}
{Polito}, V., {Reep}, J.~W., {Reeves}, K.~K., {et~al.} 2016, \apj, 816, 89

\bibitem[{{Priest} \& {Forbes}(2000)}]{Priest00}
{Priest}, E., \& {Forbes}, T. 2000, {Magnetic Reconnection}

\bibitem[{{Saint-Hilaire} {et~al.}(2008){Saint-Hilaire}, {Krucker}, \&
  {Lin}}]{Saint-Hilaire08}
{Saint-Hilaire}, P., {Krucker}, S., \& {Lin}, R.~P. 2008, \solphys, 250, 53

\bibitem[{{Schrijver} \& {Higgins}(2015)}]{Schrijver15}
{Schrijver}, C.~J., \& {Higgins}, P.~A. 2015, \solphys, 290, 2943

\bibitem[{{Scudder} \& {Karimabadi}(2013)}]{Scudder13}
{Scudder}, J.~D., \& {Karimabadi}, H. 2013, \apj, 770, 26

\bibitem[{{Scudder} \& {Olbert}(1979)}]{Scudder79}
{Scudder}, J.~D., \& {Olbert}, S. 1979, \jgr, 84, 2755

\bibitem[{{Scullion} {et~al.}(2016){Scullion}, {Rouppe van der Voort},
  {Antolin}, {Wedemeyer}, {Vissers}, {Kontar}, \& {Gallagher}}]{Scullion16}
{Scullion}, E., {Rouppe van der Voort}, L., {Antolin}, P., {et~al.} 2016, \apj,
  833, 184

\bibitem[{{Sim{\~o}es} {et~al.}(2013){Sim{\~o}es}, {Fletcher}, {Hudson}, \&
  {Russell}}]{Simoes13a}
{Sim{\~o}es}, P.~J.~A., {Fletcher}, L., {Hudson}, H.~S., \& {Russell}, A.~J.~B.
  2013, \apj, 777, 152

\bibitem[{{Sim{\~o}es} {et~al.}(2015){Sim{\~o}es}, {Graham}, \&
  {Fletcher}}]{Simoes15}
{Sim{\~o}es}, P.~J.~A., {Graham}, D.~R., \& {Fletcher}, L. 2015, \solphys, 290,
  3573

\bibitem[{{Smith} \& {Hughes}(2010)}]{Smith10}
{Smith}, R.~K., \& {Hughes}, J.~P. 2010, \apj, 718, 583

\bibitem[{{Song} {et~al.}(2015){Song}, {Chen}, {Zhang}, {Cheng}, {Wang}, {Hu},
  {Li}, \& {Wang}}]{Song15}
{Song}, H.~Q., {Chen}, Y., {Zhang}, J., {et~al.} 2015, \apjl, 808, L15

\bibitem[{{Strong}(1978)}]{Strong78}
{Strong}, K.~T. 1978, PhD thesis, , Univ.~College London, (1978)

\bibitem[{{Sun} {et~al.}(2014){Sun}, {Cheng}, \& {Ding}}]{Sun14}
{Sun}, J.~Q., {Cheng}, X., \& {Ding}, M.~D. 2014, \apj, 786, 73

\bibitem[{{Sun} {et~al.}(2013){Sun}, {Hoeksema}, {Liu}, {Aulanier}, {Su},
  {Hannah}, \& {Hock}}]{Sun13}
{Sun}, X., {Hoeksema}, J.~T., {Liu}, Y., {et~al.} 2013, \apj, 778, 139

\bibitem[{{Sweet}(1958)}]{Sweet58}
{Sweet}, P.~A. 1958, in IAU Symposium, Vol.~6, Electromagnetic Phenomena in
  Cosmical Physics, ed. B.~{Lehnert}, 123

\bibitem[{{Syntelis} {et~al.}(2016){Syntelis}, {Gontikakis}, {Patsourakos}, \&
  {Tsinganos}}]{Syntelis16}
{Syntelis}, P., {Gontikakis}, C., {Patsourakos}, S., \& {Tsinganos}, K. 2016,
  \aap, 588, A16

\bibitem[{{Tandberg-Hanssen} \& {Emslie}(1988)}]{Tandberg88}
{Tandberg-Hanssen}, E., \& {Emslie}, A.~G. 1988, {The physics of solar flares}

\bibitem[{{Testa} {et~al.}(2012){Testa}, {Drake}, \& {Landi}}]{Testa12}
{Testa}, P., {Drake}, J.~J., \& {Landi}, E. 2012, \apj, 745, 111

\bibitem[{{Vasyliunas}(1968{\natexlab{a}})}]{Vasyliunas68a}
{Vasyliunas}, V.~M. 1968{\natexlab{a}}, \jgr, 73, 2839

\bibitem[{{Vasyliunas}(1968{\natexlab{b}})}]{Vasyliunas68b}
{Vasyliunas}, V.~M. 1968{\natexlab{b}}, in Astrophysics and Space Science
  Library, Vol.~10, Physics of the Magnetosphere, ed. {R.~D.~L.~Carovillano \&
  J.~F.~McClay}, 622

\bibitem[{{Veck} {et~al.}(1984){Veck}, {Strong}, {Jordan}, {Simnett},
  {Cargill}, \& {Priest}}]{Veck84}
{Veck}, N.~J., {Strong}, K.~T., {Jordan}, C., {et~al.} 1984, \mnras, 210, 443

\bibitem[{{Veronig} {et~al.}(2010){Veronig}, {Ryb{\'a}k}, {G{\"o}m{\"o}ry},
  {Berkebile-Stoiser}, {Temmer}, {Otruba}, {Vr{\v s}nak}, {P{\"o}tzi}, \&
  {Baumgartner}}]{Veronig10}
{Veronig}, A.~M., {Ryb{\'a}k}, J., {G{\"o}m{\"o}ry}, P., {et~al.} 2010, \apj,
  719, 655

\bibitem[{{Warren} \& {Reeves}(2001)}]{Warren01}
{Warren}, H.~P., \& {Reeves}, K.~K. 2001, \apjl, 554, L103

\bibitem[{{Warren} {et~al.}(2012){Warren}, {Winebarger}, \&
  {Brooks}}]{Warren12}
{Warren}, H.~P., {Winebarger}, A.~R., \& {Brooks}, D.~H. 2012, \apj, 759, 141

\bibitem[{{Whiteford} {et~al.}(2002){Whiteford}, {Badnell}, {Ballance}, {Loch},
  {O'Mullane}, \& {Summers}}]{Whiteford02}
{Whiteford}, A.~D., {Badnell}, N.~R., {Ballance}, C.~P., {et~al.} 2002, Journal
  of Physics B Atomic Molecular Physics, 35, 3729

\bibitem[{{Whiteford} {et~al.}(2001){Whiteford}, {Badnell}, {Ballance},
  {O'Mullane}, {Summers}, \& {Thomas}}]{Whiteford01}
---. 2001, Journal of Physics B Atomic Molecular Physics, 34, 3179

\bibitem[{{Witthoeft} {et~al.}(2006){Witthoeft}, {Badnell}, {del Zanna},
  {Berrington}, \& {Pelan}}]{Witthoeft06}
{Witthoeft}, M.~C., {Badnell}, N.~R., {del Zanna}, G., {Berrington}, K.~A., \&
  {Pelan}, J.~C. 2006, \aap, 446, 361

\bibitem[{{Witthoeft} {et~al.}(2007){Witthoeft}, {Del Zanna}, \&
  {Badnell}}]{Witthoeft07}
{Witthoeft}, M.~C., {Del Zanna}, G., \& {Badnell}, N.~R. 2007, \aap, 466, 763

\bibitem[{{Woods} {et~al.}(2012){Woods}, {Eparvier}, {Hock}, {Jones},
  {Woodraska}, {Judge}, {Didkovsky}, {Lean}, {Mariska}, {Warren}, {McMullin},
  {Chamberlin}, {Berthiaume}, {Bailey}, {Fuller-Rowell}, {Sojka}, {Tobiska}, \&
  {Viereck}}]{Woods12}
{Woods}, T.~N., {Eparvier}, F.~G., {Hock}, R., {et~al.} 2012, \solphys, 275,
  115

\bibitem[{{Wright} {et~al.}(2017){Wright}, {Hannah}, {Grefenstette},
  {Glesener}, {Krucker}, {Hudson}, {Smith}, {Marsh}, {White}, \&
  {Kuhar}}]{Wright17}
{Wright}, P.~J., {Hannah}, I.~G., {Grefenstette}, B.~W., {et~al.} 2017, \apj,
  844, 132

\bibitem[{{Young} {et~al.}(2013){Young}, {Doschek}, {Warren}, \&
  {Hara}}]{Young13}
{Young}, P.~R., {Doschek}, G.~A., {Warren}, H.~P., \& {Hara}, H. 2013, \apj,
  766, 127

\bibitem[{{Zatsarinny} {et~al.}(2006){Zatsarinny}, {Gorczyca}, {Fu}, {Korista},
  {Badnell}, \& {Savin}}]{Zatsarinny06}
{Zatsarinny}, O., {Gorczyca}, T.~W., {Fu}, J., {et~al.} 2006, \aap, 447, 379

\bibitem[{{Zatsarinny} {et~al.}(2005{\natexlab{a}}){Zatsarinny}, {Gorczyca},
  {Korista}, {Fu}, {Badnell}, {Mitthumsiri}, \& {Savin}}]{Zatsarinny05a}
{Zatsarinny}, O., {Gorczyca}, T.~W., {Korista}, K.~T., {et~al.}
  2005{\natexlab{a}}, \aap, 438, 743

\bibitem[{{Zatsarinny} {et~al.}(2005{\natexlab{b}}){Zatsarinny}, {Gorczyca},
  {Korista}, {Fu}, {Badnell}, {Mitthumsiri}, \& {Savin}}]{Zatsarinny05b}
---. 2005{\natexlab{b}}, \aap, 440, 1203

\bibitem[{{Zharkova} {et~al.}(2011){Zharkova}, {Arzner}, {Benz}, {Browning},
  {Dauphin}, {Emslie}, {Fletcher}, {Kontar}, {Mann}, {Onofri}, {Petrosian},
  {Turkmani}, {Vilmer}, \& {Vlahos}}]{Zharkova11}
{Zharkova}, V.~V., {Arzner}, K., {Benz}, A.~O., {et~al.} 2011, \ssr, 156

\bibitem[{{Zweibel} \& {Yamada}(2009)}]{Zweibel09}
{Zweibel}, E.~G., \& {Yamada}, M. 2009, \araa, 47, 291

\end{thebibliography}

\end{document}